\documentclass[12pt]{article}
\usepackage{amsmath}
\usepackage{amsfonts}
\usepackage{amssymb}
\usepackage{srcltx}

\newtheorem{theorem}{Theorem}
\newtheorem{proposition}[theorem]{Proposition}
\newtheorem{lemma}[theorem]{Lemma}

\newtheorem{definition}[theorem]{Definition}

\newtheorem{remark}[theorem]{Remark}
\newtheorem{example}[theorem]{Example}
\newcommand{\ptl}{\partial}
\begin{document}
\hfill\today\\
\noindent
{\Large {\bf Equivalence of formulations of the MKP hierarchy and its polynomial tau-functions }}
\vskip 9 mm
\begin{center}
\begin{minipage}[t]{70mm}
{\bf Victor Kac}\\
\\
Department of Mathematics,\\
Massachusetts Institute of Technology,\\
Cambridge, Massachusetts 02139, U.S.A\\
e-mail: kac@math.mit.edu\\
\end{minipage}\qquad
\begin{minipage}[t]{70mm}
{\bf Johan van de Leur}\\
\\
Mathematical Institute,\\
Utrecht University,\\
P.O. Box 80010, 3508 TA Utrecht,\\
The Netherlands\\
e-mail: J.W.vandeLeur@uu.nl
\end{minipage}
\end{center}
\section{Introduction}
The KP hierarchy was introduced by Sato in his seminal paper \cite{S} as the hierarchy of evolution equations of Lax type
\[ \frac{dL}{dt_n} = [ (L^n)_+, L  ], \ n = 1, 2, \ldots,  \]
on the pseudodifferential operator $ L = \ptl + u_1 \ptl^{-1} + u_2 \ptl^{-2} + \ldots,  $ where $ \ptl = \frac{\ptl}{\ptl t_1} $ and $ + $ stands for the differential part. He also introduced the associated wave functions and the tau-function, and discussed reductions of the KP hierarchy. His ideas have  been subsequently developed by his school in a series of papers, which were reviewed in \cite{JM} and \cite{DJKM}. 

In the review \cite{JM} Jimbo and Miwa also introduced the
modified KP hierarchy (MKP hierarchy), 
as a set of bilinear equations on the tau-functions $\tau_\ell$, $\ell\in\mathbb{Z}$, see  \cite{JM}, eq.$(2.4)_{\ell,\ell'}$, each $\tau_\ell$  being a 
tau-function of the KP hierarchy. It was subsequently shown in \cite{KP} that these equations arise naturally from the fermionic formulation of the 
MKP hierarchy and the boson-fermion correspondence. This implies that the MKP tau-functions $(\ldots,\tau_{\ell-1},\tau_\ell, \tau_{\ell+1},\ldots)$ 
are naturally parameterized by the infinite-dimensional flag manifold (\cite{KP}, Corollary 8.1), in analogy with the famous observation of Sato \cite{S} 
that tau-functions of the KP hierarchy are parametrized by the infinite-dimensional Grassmann manifold.  Note that the tau-functions of the discrete KP hierarchy, studied in \cite{AM}, are precisely those, satisfying the Jimbo-Miwa equations from \cite{JM}. 

On the other hand, Dickey proposed a Lax type formulation of the MKP hierarchy in \cite{D} (see also \cite{Dbook}), 
which is an extension of the Sato formulation of KP.
The first result of the present paper is the equivalence of Jimbo-Miwa's tau-function formulation and Dickey's Lax type formulation 
of the MKP hierarchy (Theorem \ref{theorem2} in Section \ref{S4}), in analogy with the well developed theory of the KP hierarchy (see e.g. \cite{JM}, \cite{DJKM}). 
Similar equivalences are established for the discrete KP hierarchy in
\cite{AM}. The vertex operator construction of the Lie agebra $g\ell_{\infty}$ provides solutions to the
tau-function formulation of the MKP hierarchy \cite{KP}, hence to the Lax type formulation of it.
Similar solutions have been constructed in \cite{AM} for the discrete KP hierarchy.

In Section \ref{Seigen}
we give
eigenfunction formulations of the MKP hierarchy, closely related to the work \cite{HL}. As a byproduct, we find in Section \ref{S7} an astonishingly simple explicit description of all polynomial tau-functions
of the KP and the MKP hierarchies (Theorem \ref{T7}).
Of course, it is a well-known result of Sato \cite{S} that all Schur polynomials are tau-functions of the KP hierarchy. We show that, moreover, all polynomial tau-functions of the KP hierarchy can be
obtained from Schur polynomials by certain shifts of arguments.

We discuss in Section \ref{S5} the reductions of the MKP hierarchy to the modified $n$-KdV hierarchies for each integer $n\geq 2$,
the $n=2$ case being the classical modified KdV hierarchy (cf. \cite{D}).
Finally, in Section \ref{S8} we find all polynomial tau-functions for the $n$-KdV hierarchy, and (implicitly) for the modified $n$-KdV hierarchy. This was known only for $n=2$ \cite{KP}.

\section{The  fermionic formulation of MKP}
\label{S2}
Recall the semi-infinite wedge representation \cite{KP}, \cite {KL}.
Consider the infinite  matrix group
$GL_{\infty}$, consisting of all complex matrices  $G= (g_{ij})_{i,j \in {\Bbb Z}}$ which are 
 invertible and all but a finite number of $g_{ij} -
\delta_{ij}$ are $0$.
It acts naturally on the vector space 
${\Bbb C}^{\infty} = \bigoplus_{j \in {\Bbb Z}} {\Bbb
C} e_{j}$  (via the usual formula
$E_{ij} (e_{k}) = \delta_{jk} e_{i}$).

 The semi-infinite wedge space $F =
\Lambda^{\frac{1}{2}\infty} {\Bbb C}^{\infty}$ is the vector space
with a basis consisting of all semi-infinite monomials of the form
$e_{i_{1}} \wedge e_{i_{2}} \wedge e_{i_{3}} \ldots$, where $i_{1} >
i_{2} > i_{3} > \ldots$ and $i_{\ell +1} = i_{\ell} -1$ for $\ell >>
0$.  One defines the  representation $R$ of $GL_{\infty}$  on $F$ by
$$
R(G) (e_{i_{1}} \wedge e_{i_{2}} \wedge e_{i_{3}} \wedge \cdots) = G
e_{i_{1}} \wedge G e_{i_{2}} \wedge Ge_{i_{3}} \wedge \cdots , 
$$
and apply linearity and anticommutativity of  the wedge product $\wedge$.

The corresponding representation $r$ of the Lie algebra
$g\ell_{\infty}$
of $GL_\infty$ can be described
in  terms of a Clifford algebra.  Define the wedging and contracting
operators $\psi^{+}_{j}$ and $\psi^{-}_{j}\ \ (j \in {\Bbb Z} +
\frac{1}{2})$ on $F$ by
$$\begin{aligned}
&\psi^{+}_{j} (e_{i_{1}} \wedge e_{i_{2}} \wedge \cdots ) =  
e_{-j+\frac12}\wedge e_{i_{1}} \wedge e_{i_{2}} \cdots, \\
&\
\psi^{-}_{j} (e_{i_{1}} \wedge e_{i_{2}} \wedge \cdots ) = \begin{cases} 0
&\text{if}\ j-\frac12 \neq i_{s}\ \text{for all}\ s \\
(-1)^{s+1} e_{i_{1}} \wedge e_{i_{2}} \wedge \cdots \wedge
e_{i_{s-1}} \wedge e_{i_{s+1}} \wedge \cdots &\text{if}\ j = i_{s}-\frac12.
\end{cases}
\end{aligned}
$$
These operators satisfy the relations
$(i,j \in {\Bbb Z}+\frac{1}{2}, \lambda ,\mu = +,-)$:
$$\psi^{\lambda}_{i} \psi^{\mu}_{j} + \psi^{\mu}_{j}
\psi^{\lambda}_{i} = \delta_{\lambda ,-\mu} \delta_{i,-j}, $$
hence they generate a Clifford algebra, which we denote by ${\cal C}\ell$.
Introduce the following elements of $F$ $(m \in {\Bbb Z})$:
\begin{equation}
\label{vac}
|m\rangle = e_{m } \wedge e_{m-1 } \wedge
e_{m-2} \wedge \cdots .
\end{equation}
It is clear that $F$ is an irreducible ${\cal C}\ell$-module such that
$$\psi^{\pm}_{j} |0\rangle = 0 \ \text{for}\ j > 0 . $$
The representation $r$ of $g\ell_\infty$ in $F$, corresponding to the representation $R$ of $GL_\infty$, is given by the
 formula
$r(E_{ij}) = \psi^{+}_{-i+\frac12} \psi^{-}_{j-\frac12}. $
Define the {\it charge decomposition}
$$F = \bigoplus_{m \in {\Bbb Z}} F^{(m)}, \quad \text{
where }
\text{charge}(|m\rangle ) = m\ \text{and charge} (\psi^{\pm}_{j}) =
\pm 1. $$
The space $F^{(m)}$ is an irreducible highest weight
$g\ell_{\infty}$-module, with highest weight vector
$|m\rangle$:
$$
r(E_{ij})|m\rangle = 0 \ \text{for}\ i < j,\quad
r(E_{ii})|m\rangle = 0\  (\text{resp.}\ = |m\rangle ) \ \text{if}\ i > m\
(\text{resp. if}\ i \le m).
$$
Let
$${\cal O}_m
= R(GL_{\infty})|m\rangle \subset F^{(m)}$$ be the $GL_{\infty}$-orbit
of the highest weight vector $|m\rangle$.

\begin{theorem}{\rm  (\cite{KP}, Theorem 5.1)}
Let $I$ be a non-empty finite subset of $\mathbb{Z}$ and let $f=\oplus_{m\in I}f_m\in\oplus_{m\in I}F^{(m)}$ be such that all $f_m\not = 0$.
Then $f\in \oplus_{m\in I} {\cal O}_m$ if and only if for all $k,\ell\in I$, such that $k\ge \ell$, one has
\begin{equation}\
\label{modKP}
\sum_{i\in\mathbb{Z}+
\frac{1}{2}} \psi_i^+ f_k\otimes \psi_{-i}^- f_\ell =0\, .
\end{equation}
\end{theorem}

Equation (\ref{modKP}) is called the $(k-\ell)$-th modified KP hierarchy in the fermionic picture. The 0-th modified KP is the KP hierarchy.
The collection of all such equations $k,\ell\in\mathbb{Z}$ with $k\ge \ell$ is called the (full) MKP hierachy in the fermionic picture.

\section{The  bosonic formulation of MKP}
\label{S3}
Define the fermionic fields by 
$
\psi^\pm(z)=\sum_{i\in\mathbb{Z}+\frac12}\psi_i^\pm z^{-i-\frac12}
$
and the bosonic field
$\alpha(z)=\sum_{n\in\mathbb{Z}}\alpha_n z^{-n-1}=:\psi^+(z)\psi^-(z):$.
Then there exists a unique vector space isomorphism, called the boson-fermion correspondence,
$ \sigma:F \to B=\mathbb{C}[q,q^{-1}]\otimes \mathbb{C}[t_1,t_2,\ldots]$ 
such that $\sigma (|m\rangle)=q^m$, $\sigma \alpha_n\sigma^{-1}=\frac{\partial}{\partial t_n}$, 
$\sigma \alpha_{-n}\sigma^{-1}=n t_n$, for $n>0$ and  $\sigma \alpha_0\sigma^{-1}=q\frac{\partial}{\partial q}$. Moreover, one has
\begin{equation}
\label{2}
\sigma \psi^\pm(z)\sigma^{-1}= q^{\pm 1}z ^{\pm q\frac{\partial}{\partial q}}\exp\left(\pm \sum_{k=1}^\infty t_kz^k\right)
\exp\left(\mp \sum_{k=1}^\infty \frac{\partial}{\partial  t_k}\frac{ z^{-k}}{k}\right).
\end{equation}
For $f_m\in{\cal O}_m\cup\{ 0\}$  we write: $\sigma (f_m)= \tau_m(t) q^m$, where $t=(t_1,t_2,\ldots)$. Such a $\tau_m$ is called a tau-function.
Under the isomorphism $\sigma$ we can rewrite (\ref{modKP}), using (\ref{2}), to obtain a Hirota bilinear identity for tau-functions\\
\ 
\\
{\bf The  first formulation of the MKP hierarchy:}
\\
{\it 
Let 
$[z]=(z,\frac{z^2}2,\frac{z^3}3,\ldots)$, $y=(y_1,y_2,\ldots)$, and ${\rm Res}\,  \sum_i f_iz^i dz=f_{-1}$, then 
\begin{equation}
\label{modKP2}
{\rm Res}\,   z^{k-\ell}\tau_k(t-[z^{-1}])\tau_\ell(y+[z^{-1}]) \exp\left(\sum_{i=1}^\infty (t_i-y_i)z^i\right)dz=0,\qquad k\ge \ell.
\end{equation}
}
The equations (\ref{modKP2})
first appeared in \cite{JM}, $(2.4)_{l,l'}$.

Divide (\ref{modKP2}) by $\tau_k(t)\tau_\ell(y)$ and introduce the wave functions $w_m^+$ and adjoint wave function $w_m^-$ ($m\in\mathbb{Z})$ by
\begin{equation}
\label{wave}
\begin{aligned}
w_m^\pm(t,z)=&q^{\mp 1}
\frac{\sigma\left( \psi^\pm(z)f_m\right)}{\sigma(f_m)}\\
=&
{z^{\pm m}}\frac{\tau_m(t\mp [z^{-1}])}{\tau_m(t)} e^{\pm t\cdot z}\, .
\end{aligned}
\end{equation}
Here and thereafter  we use the shorthand notation $$t\cdot z =\sum_{i=1}^\infty t_iz^i\, .$$
Then (\ref{modKP2}) becomes
\begin{equation}
\label{modKP3}
{\rm Res}\,  w^+_k(t,z)w^-_\ell (y,z)dz =0, \qquad k\ge \ell.
\end{equation}

\section{The Lax type formulation of MKP}
\label{S4}
We now want to express the wave functions in terms of formal pseudodifferential operators in 
$\partial=\frac{\partial}{\partial t_1}$. A formal pseudodifferential operator is an expression of the form 
\[
P(t,\partial ) = \sum_{j \leq N} P_{j}(t)\partial^{j}\, ,
\]
where the $P_j(t)$ are functions in $t$, infinitely differentiable in $t_1$.
The differential part of $P(t,\partial )$ is
$P_{+}(t,\partial ) := \sum^{N}_{j=0} P_{j}(t) \partial^{j}$, 
and $P_{-} := P-P_{+}$.  
These operators form an associative algebra with multiplication $\circ$, defined by ($k,\ell\in\mathbb{Z}$)
\[
A(t)\partial^k\circ B(t)\partial^\ell=\sum_{i=0}^\infty \begin{pmatrix}k\\ i\end{pmatrix}\frac{\partial^i A(t)}{\partial t_1^i}B(t)\partial^{k+\ell-i}\,. 
\]
The formal adjoint of $P(t,\partial)$ is defined by the
following formula:
$$(\sum_{j} P_{j}(t)\partial^{j})^{*} = \sum_{j} (-\partial )^{j}
\circ P_{j}(t). $$
The residue of $P(t,\partial)$ is ${\rm Res}_\partial P(t,\partial):=P_{-1}(t)$.

Let
\begin{equation}
\label{wwave}
P_m^\pm(t,\pm z)=\frac{\tau_m(t\mp [z^{-1}])}{\tau_m(t)}=1\pm p_1^\pm(t)z^{-1}+p_2^\pm(t)z^{-2}\pm\cdots,
\end{equation}
so that 
\begin{equation}
\label{wwwave}
\begin{aligned}
w^\pm_m(t,z)&=P_m^\pm(t,\pm z)z^{\pm m}e^{\pm t\cdot z}=P_m^\pm(t,\partial)\circ(\pm\partial)^{\pm m}(e^{\pm t\cdot z})\\
&=P_m^\pm(t,\partial)\circ \partial^{\pm m}\circ\exp\left( \pm \sum_{i=2}^\infty t_i(\pm \partial)^i\right) (e^{\pm t_1z})
\, .
\end{aligned}
\end{equation}
Then (\ref{modKP2}) is equivalent to
\begin{equation}
\label{modKP4}
{\rm Res}\, 
P_k^+(t, z)z^{ k}e^{ t\cdot z}
P_\ell^-(y,- z)z^{ -\ell}e^{- y\cdot z}
 dz=0.
\end{equation}

The following lemma is crucial. It involves only the first variable $t_1$. When
we use it, the variables $t_2, t_3,\ldots$ are seen as extra parameters.
\begin{lemma}
\label{L1}
 {\rm (\cite{KL}, Lemma 4.1)} Let $P(t_1,\partial)$ and $Q(t_1,\partial)$ be two formal pseudo-differential operators, then
\[
{\rm Res}\,  P(t_1,z)e^{t_1z} Q(y_1, -z)e^{-y_1z}dz ={\rm Res}\, _\partial P(t_1,\partial)\circ Q(t_1,\partial)^*\circ e^{u\partial}|_{u=t_1-y_1}\, .
\]
\end{lemma}
\indent

Applying the lemma to the bilinear identity (\ref{modKP3}), while using  the expression  (\ref{wwwave}) for the wave functions, one deduces
\begin{equation}
\label{PP-}
P_k^-(t, \partial)^*=P_k^+(t, \partial)^{-1},\quad \left(P_k^+(t, \partial)\circ \partial^{(k-\ell)}\circ P_\ell^+(t, \partial)^{-1}\right)_-=0.
\end{equation}
We obtain the  Sato-Wilson equation
\begin{equation}
\label{Sato}
\frac{\partial P_k^+(t, \partial)}
{\partial t_j}=\left(P_k^+(t, \partial)\circ \partial^j\circ P_k^+(t, \partial)^{-1}\right)_-\circ P_k^+(t, \partial)\, ,
\end{equation}
by differentiating (\ref{modKP3}) by $t_j$, using the first equation of (\ref{PP-}) and then applying Lemma \ref{L1}
 (see  e.g. \cite{KL}, proof of  Lemma 4.2) .

Introduce the Lax operator $L_k$ by dressing $\partial$ by (the dressing operator) $P_k^+$:
\begin{equation}
\label{Laxop}
L_k=L_k(t,\partial)=P_k^+(t, \partial)\circ\partial\circ P_k^+(t, \partial)^{-1}.
\end{equation}
Differentiate {(\ref{wwwave}) by $t_j$ and apply the Sato-Wilson equation (\ref{Sato}). This gives the following linear equation (= linear problem) for the wave function $w_k^+$ ($k\in \mathbb{Z}$):
\begin{equation}
\label{lin}
L_k w^+_k(t,z)=zw^+_k(t,z),\quad \frac{\partial w^+_k(t,z)}{\partial t_j}= \left( L_k^j\right)_+w^+_k(t,z)
\end{equation}
and the adjoint wave function $w_k^-$:
\begin{equation}
\label{linad}
L_k^* w^-_k(t,z)=zw^-_k(t,z),\quad \frac{\partial w^-_k(t,z)}{\partial t_j}= -\left( L_k^j\right)^*_+w^-_k(t,z)\, .
\end{equation}

{}From  (\ref{Sato}) it is easy to deduce the  Lax equations on $L_k$ (see e.g. \cite{KL}, Lemma 4.3):
\begin{equation}
\label{Lax}
\frac{\partial L_k}{\partial t_j}=\left[ (L_k^j)_+, L_k\right]\, ,\quad j=1,2,\ldots,
\end{equation}
which are the compatibility conditions of the linear problem (\ref{lin}).
{}From (\ref{wwave}) we find that 
\[
P_k^+(t, \partial)= 1-\partial (\log \tau_k(t))\partial^{-1}+\cdots,
\]
hence  the second equation of (\ref{PP-}) 
for $k=\ell+1$ gives that 
\[
P_{\ell+1}^+(t, \partial)\circ \partial\circ P_\ell^+(t, \partial)^{-1}=\left(P_{\ell+1}^+(t, \partial)\circ\partial\circ P_\ell^+(t, \partial)^{-1}\right)_+=\partial +\partial(\log(\tau_\ell(t))-\partial(\log(\tau_{\ell+1}(t))\, ,
\]
and hence 
\begin{equation}
\label {Pll}
P_{\ell+1}^+(t, \partial)\partial =(\partial +v_\ell(t))\circ P_{\ell}^+(t, \partial), \quad\text{where }v_\ell(t)= \partial \left(\log \frac{\tau_\ell(t)}{\tau_{\ell+1}(t)}\right).
\end{equation}
This leads to another  formulation of MKP, which was suggested by Dickey \cite{D}, \cite{Dbook}:
\\
\ 
\\
{\bf The second  formulation of the MKP hierarchy: }
\\
\\
{\it Let $U=\mathbb{C}[u_i^{(n)}, v_j^{(n)}| i\in\mathbb{Z}_{\ge 1},j\in\mathbb{Z}, n\in\mathbb{Z}_{\ge 0}]$ be the algebra of differential polynomials in 
$u_i$ and $v_j$, where $\partial u_i^{(n)}=u_i^{(n+1)}$, $\partial v_j^{(n)}=v_j^{(n+1)}$.
Let $L_0(\partial )=\partial +u_1(t)\partial^{-1} +u_2(t)\partial^{-2} \ldots \in U((\partial^{-1}))$ be a pseudo-differential operator. Then  the MKP hierarchy is the following system of evolution equations in $U$ ($j\in\mathbb{Z}_{\ge 1}$,  $ i\in\mathbb{Z}$):
\begin{equation}
\label{Dickey}
\frac{\partial L_0(\partial)}{\partial t_j} =[(L_0(\partial)^j)_+,L_0(\partial)], \quad \frac{\partial v_i}{\partial t_j}=\left(L_{i+1}(\partial)^j\right)_+\circ(\partial+v_i)-(\partial+v_i)\circ\left(L_{i}(\partial)^j\right)_+ ,
\end{equation}
where $L_i(\partial)$ and $L_{-i}(\partial)$, for $i>0$, are defined by
\begin{equation}
\label{LLL}
L_i(\partial)=(\partial+ v_{i-1})\circ L_{i-1}(\partial)\circ(\partial+ v_{i-1})^{-1},\qquad
L_{-i}(\partial)=(\partial+ v_{-i})^{-1}\circ L_{1-i}(\partial)\circ (\partial+ v_{-i})\, .
\end{equation}
}
\begin{theorem}\label{theorem2}
The first and the second formulation of MKP are equivalent.
\end{theorem}

\noindent {\bf Proof.} 
To prove that the first formulation implies the second,
first note that, using the first formula of (\ref{Pll}), one indeed gets that for $\ell>0$:
\begin{equation}
\label{*}
\begin{aligned}
L_{\ell}&=P^+_{\ell}\circ\partial\circ (P^{+}_{\ell})^{-1}=(\partial+v_{\ell-1})\circ P^+_{\ell-1}\circ \partial (P^+_{\ell-1})^{-1}\circ(\partial+v_{\ell-1})^{-1}\\
&=
(\partial+v_{\ell-1})\circ L_{\ell-1}\circ (\partial+v_{\ell-1})^{-1}\quad\text{and}\\
L_{-\ell}&=P^+_{-\ell}\circ \partial \circ (P^+_{-\ell})^{-1}=(\partial+v_{-\ell})^{-1}\circ P^+_{1-\ell}\circ \partial\circ (P^+_{1-\ell})^{-1}\circ (\partial+v_{-\ell})\\
&=(\partial+v_{-\ell})^{-1}\circ L_{1-\ell}\circ (\partial+v_{-\ell})\, .
\end{aligned}
\end{equation}
Secondly, we show that the second equation of (\ref{Dickey}) holds. This follows from the Sato-Wilson equation (\ref{Sato}).
Indeed, 
\[
\begin{aligned}
\frac{\partial P_{\ell+1}^+(t, \partial)}{\partial t_j}=&-\left(L_{\ell +1}(t, \partial)^j\right)_-\circ (\partial+v_\ell(t))\circ P_\ell^+(t, \partial)\\
=&
\frac{\partial v_{\ell}(t)}{\partial t_j}
P_\ell^+(t, \partial)
-(\partial+v_\ell(t))\circ \left(L_{\ell }(t, \partial)^j\right)_-\circ P_\ell^+(t, \partial)\, ,
\end{aligned}
\]
we deduce that 
\[
\begin{aligned}
\frac{\partial v_{\ell}(t)}{\partial t_j}=&
-\left(L_{\ell +1}(t, \partial)^j\right)_-\circ (\partial+v_\ell(t))+(\partial+v_\ell(t))\circ \left(L_{\ell }(t, \partial)^j\right)_-\\
=&-L_{\ell +1}(t, \partial)^j\circ (\partial+v_\ell(t))+\left( L_{\ell +1}(t, \partial)^j\right)_+\circ (\partial+v_\ell(t))\\
&\qquad\qquad
+(\partial+v_\ell(t))\circ L_{\ell }(t, \partial)^j-
(\partial+v_\ell(t))\circ \left(L_{\ell }(t, \partial)^j\right)_+\\
=&\left( L_{\ell +1}(t, \partial)^j\right)_+\circ (\partial+v_\ell(t))
-
(\partial+v_\ell(t))\circ \left(L_{\ell }(t, \partial)^j\right)_+\, .
\end{aligned}
\]
Here we have used that  $L_{\ell +1}(t, \partial)^j\circ (\partial+v_\ell(t))=(\partial+v_\ell(t))\circ L_{\ell }(t, \partial)^j$.

To prove the converse, we use a result of Shiota  \cite{Sh}, the Claim of Section 1.2.
He shows that  if $L_0$ satisfies the Lax equation (\ref{Lax}),  then $w_0^+(t,z)$  is uniquely determined by the linear problem (\ref{lin}), up to multiplication by 
elements of the form $1+\sum_{i>0}a_iz^{-i}$, with $a_i\in\mathbb{C}$ or rather $P_0(t,\partial)=1+\sum_{i>0}w_i(t)\partial^{-i}$ is a unique solution up to 
multiplication from the right by elements of the form $1+\sum_{i>0}a_i\partial^{-i}$, with $a_i\in\mathbb{C}$, of the equations
\[
L_0\circ P_0^+(t,\partial)=P_0^+(t,\partial)\circ \partial,\quad \frac{\partial P_0^+(t,\partial)}{\partial t_j}=P_0^+(t,\partial)\circ\partial^j-\left(L_0^j\right)_+ \circ P_0^+(t,\partial)\, .
\]
Hence,  
$w_0^+(t)=P_0^+(t,\partial)e^{t\cdot z}$ satisfies (\ref{lin}) and thus   is a wave function for $L_0$, so that  $w_0^-(t)=(P_0^+(t,\partial))^{*-1}e^{-t\cdot z}$ is the adjoint wave function. For $i>0$ let  
\[
\begin{aligned}
P_i^+&=(\partial +v_{i-1})\circ(\partial+v_{i-2})\circ \cdots\circ  (\partial+v_0)\circ P^+_0, \\
P_{-i}^+&=(\partial +v_{-i})^{-1}\circ (\partial+v_{1-i})^{-1}\circ\cdots\circ  (\partial+v_{-1})^{-1}\circ P^+_0\, ,
\end{aligned}
\]
and construct all other (adjoint) wave functions via
\begin{equation}
\label{**}
\begin{aligned}
w_i^+=&(\partial+v_{i-1})(w_{i-1}^+),
\quad
&w_i^-= (\partial+v_{i-1})^{*-1}(w_{i-1}^-),
\\
w_{-i}^+=& (\partial+v_{-i})^{-1}(w_{1-i}^+),
\quad
&w_{-i}^-=(\partial+v_{-i})^{*}(w_{1-i}^-)\, .
\end{aligned}
\end{equation} 
By(\ref{Dickey}) and (\ref{LLL}) these (adjoint) wave functions satisfy the linear problem (\ref{lin}).
In order to show that the bilinear identity holds for the wave functions, we first prove that
\begin{equation}
\label{inproof1}
\left(\partial^jP_{k}^+(t,\partial)P_\ell^{-*}(t,\partial)\right)_-=0\quad \text{for all }\ k\ge \ell,\ j\ge 0.
\end{equation}
We show this for $k\ge 0$ and $\ell<0$ (all other cases are obvious):
\[
\begin{aligned}
\partial^j\circ P_{k}^+P_\ell^{-*}=&\partial^j\circ (\partial +v_{k-1})\circ \cdots\circ  (\partial+v_0)\circ P^+_0\circ (P^+_0)^{-1}\circ (\partial+v_{-1})\circ \cdots\circ (\partial+v_{\ell})\\
=&\partial^j\circ (\partial +v_{k-1})\circ (\partial+v_{k-2})\circ \cdots \circ (\partial+v_{\ell}).
\end{aligned}
\]
Using  Lemma \ref{L1}, we deduce from (\ref{inproof1}) that
\[
{\rm Res}\,  \frac{\partial^j w_k^+(s_1, t_2, t_3\cdots,z)}{\partial s_1^j}w_\ell^-(t_1,t_2,t_3,\ldots,z)dz=0.
\]
The second formula of (\ref{lin}) implies that
\[
{\rm Res}\,  \frac{\partial^{j_1+j_2+\cdots+ j_n} w_k^+(s_1, t_2, t_3\cdots,z)}{\partial s_1^{j_1}\partial t_2^{j_2}\cdots\partial t_n^{j_n}}w_\ell^-(t_1,t_2,t_3,\ldots,z)dz=0.
\]
Using Taylor's formula we obtain the bilinear identity (\ref{modKP3}) for the wave function.
The tau-functions  $\tau_i$ are then obtained up to a scalar factor by the formula (see e.g. \cite{KL} eq. (111), which  is a direct consequence of (\ref{wwave})):
\[
\frac{\partial \log \tau_i(t)}{\partial t_j}={\rm Res}\,  z^j\left(\frac{\partial}{\partial z}-\sum_{k>0} z^{-k-1}\frac{\partial }{\partial t_k}\right) P^+_i(t,z)\, .
\]
Hence, multiplying (\ref{modKP3}) by $\tau_k(t)\tau_\ell (y)$, we obtain the bilinear identities (\ref{modKP2}) for the tau-functions, which is the first formulation of MKP.
Thus the two formulations are equivalent.\hfill$\square$
\\
\
\\
\indent The $v_j$ are expressed in terms of the tau-functions via the second formula of (\ref{Pll}). Using (\ref{wwave}), we see that 
\[
P_0^\pm (t,\partial)= \sum_{i,j=0}^\infty \frac{S_i(\mp D )\tau_0}{\tau_0}\partial^{-i},\ \text{where }
\sum_{i=0}^\infty S_i(D)z^i=\exp\left(\sum_{k=1}^\infty \frac{z^k}k\frac{\partial}{\partial t_k}\right).
\]
This and the fact that $L_0$ is given by (\ref{Laxop}), gives that the  $u_i$ can be calculated by the
following formula
\[
L_0(t, \partial)=\sum_{i,j=0}^\infty \frac{S_i(-D)\tau_0}{\tau_0}\partial ^{1-i-j} \circ \frac{S_j(D)\tau_0}{\tau_0}\, .
\]
\begin{remark}
\label{remark1}
Dickey shows that  all flows $\frac{\partial}{\partial t_k}$, defined by (\ref{Dickey}), commute (\cite{D}, Proposition 2.3). Hence (\ref{Dickey}) is an integrable
system of compatible evolution equations in $U$.
\end{remark}
\begin{remark}
\label{remark2}
The differential algebra $U $ carries an automorphism $ S $ (commuting with $ \partial  $), defined by
\[ S(v_j) = v_{j+1}, \quad S(L) = (\partial + v) \circ L \circ(\partial + v)^{-1}. \]
The MKP hierarchy can be understood as the following system of partial differential-difference equations ($j=1,2,...$)
\[ \begin{cases}
\frac{dL}{dt_j} = [(L^j)_+, L ] \\
\frac{dv}{dt_j} = (S(L)^j)_+\circ(\partial + v) - (\partial + v) \circ(L^j)_+ \, . \\
\end{cases} \] 
Here $ L = \partial + u_1 \partial^{-1} + u_2 \partial^{-2} + \ldots  $ and $ v = v_0. $
\end{remark}

\section{Eigenfunction formulation of MKP}
\label{Seigen}

There is yet another formulation of MKP. It is given in terms of eigenfunctions and adjoint eigenfunctions of the Lax operators $L_k$. 

\begin{definition}
\label{Deigen} 
Let $L=L(t, \partial)$ be a pseudodifferential operator with coefficients in $\mathbb{C}(t_1,t_2,\ldots)$, where $\partial=\frac{\partial}{\partial t_1}$.  An element $\phi\in \mathbb{C}(t_1,t_2,\ldots)$
is called an eigenfunction (resp. adjoint eigenfunction) for $L$ if
\begin{equation}
\label{eigen}
\frac{\partial \phi(t)}{\partial t_n}=\left(L^n\right)_+(\phi(t))
\quad\left(\mbox{resp. }
\frac{\partial \phi (t)}{\partial t_n}=-\left(L^{n}\right)^*_+(\phi(t))\right),\, n=1,2,\cdots \,.
\end{equation}
\end{definition}
\begin{example}
\label{EX}
Let $L=L(t,\partial)$  be a pseudodifferential operator and $w^+(t,z)$ (resp $w^-(t,z)$) satisfy
\[
 \frac{\partial w^+(t,z)}{\partial t_j}= \left( L^j\right)_+w^+(t,z),\ 
\left(\mbox{resp. }
 \frac{\partial w^-(t,z)}{\partial t_j}= -\left( L^j\right)^*_+w^-(t,z)\right)\, ,
\]
cf. (\ref{lin}) and (\ref{linad}). Then for each $f(z)\in \mathbb{C}((z^{-1}))$
 the functions 
\begin{equation}
\label{qwave}
q^\pm_f(t)= {\rm Res} f(z)w^\pm (t,z) dz\, ,
\end{equation}
are eigenfunctions (taking $+$) and adjoint eigenfunctions (taking $-$) for $L$.
In particular if $L=\partial$, then
$$
q^\pm_f(t)= {\rm Res} f(z)  e^{\pm t \cdot z} dz\,  ,
$$
are its (adjoint) eigenfunctions.
\end{example}
These (adjoint) eigenfunctions were used by Matveev and Salle \cite{MS} to construct new solutions of  the KP equation from old ones. In fact we will prove later the following 
\begin{proposition}
\label{Matv}
If $\tau(t)$ is a tau-function, satisfying (\ref{modKP2}) for $k=\ell$, and $L=P^+\circ \partial\circ  (P^{+})^{-1}$ is  the corresponding Lax operator, where $P^+$ is given by (\ref{wwave}), then $\phi^\pm(t)\tau(t)$ is again a tau-function, provided that  $\phi^\pm(t)$  is an (adjoint) eigenfunction for $L$.
\end{proposition}

We will show (see also  \cite{HL}) that $\tau(t)$ and $\phi^\pm(t)\tau(t)$ satisfy the 1st modified KP hierarchy (\ref{modKP2}) for $k-\ell=1$.
The converse of this statement  also holds, namely we have 
\begin{proposition}
\label{prop6}
Let  $\tau_k(t)$ and $\tau_{k+1}(t)$ be KP tau-functions that satisfy  (\ref{modKP2}) for $k-\ell=1$. Then their ratio  $\phi_k(t)=\frac{\tau_{k+1}(t)}{\tau_k(t)}$
 is an eigenfunction for $L_k=P^+_k\partial P^{+-1}_k$ and
$\frac1{\phi_k(t)}$ is an adjoint eigenfunction for  $ L_{k+1} =P^+_{k+1}\partial P^{+-1}_{k+1}$, where $P_m^+$ is given by (\ref{wwave}).
\end{proposition}
\noindent {\bf Proof.} 
The tau-function formulation of the $1$-st MKP hierarchy, i.e. (\ref{modKP2}) for  $k-\ell=1$
is equivalent to (see e.g. \cite{KL2}, Theorem 2.3 (c),  for $l=1$).
\begin{equation}
\label{mod1}
{\rm Res}\,   z^{-1}\tau_k(t-[z^{-1}])\tau_{k+1}(y+[z^{-1}]) \exp\left(\sum_{i=1}^\infty (t_i-y_i)z^i\right)dz=\tau_{k+1}(t)\tau_k(y).
\end{equation}
Divide  equation (\ref{mod1})   by $\tau_{k+1}(t) \tau_k(y)$, to obtain:
\begin{equation}
\label{mod2}
{\rm Res}\,   \phi_k(t)^{-1}w^+_k(t,z)\phi_k(y)w^-_{k+1}(y,z) dz=1\, .
\end{equation}
Differentiate this equation by  $t_n$ and then multiply by $\phi_k(t)$, to obtain
\[
{\rm Res}\,   \left(
-\frac{\partial \phi_k(t)}{\partial t_n}\phi_k(t)^{-1}w^+_k(t,z)+  \left(L_k^n\right)_+(w^+_k(t,z))\right)
\phi_k(y)w^-_{k+1}(y,z) dz=0.
\]
Using Lemma \ref{L1}, (\ref{wwave}), (\ref{wwwave}) and the fact that 
\[
w_{k+1}^-(y,z)=\frac1{\phi_k(y)}(-\partial)^{-1}\circ\phi_k(y)\left(P_k^+(y, \partial)\right)^{*-1}e^{-\sum_i y_iz^i}\, ,
\]
we obtain 
\[
\left(-\frac{\partial \phi_k(t)}{\partial t_n}\phi_k(t)^{-1}P_k^+(t) \circ P_k^+(t)^{-1}\circ\phi_k(t)\partial^{-1}+ (L_k^n)_+\circ P_k^+(t) \circ P_k^+(t)^{-1}\circ\phi_k(t)\partial^{-1}\right)_-=0\, .
\]
Taking the residue of this expression (i.e. the coefficient of $\partial^{-1}$) gives  equation (\ref{eigen}).
The second formula can be also obtained from (\ref{mod2}) in almost the same way, but now one has to differentiate this equation by $y_1$ and  continue in a similar manner.
\hfill$\square$
\\
\
\\
One also has
\begin{proposition}
\label{Pkk}
Let $\phi_k(t)$ be as in the previous Proposition and 
let $w_k^+(t,z)=P^+_k(t,z)z^ke^{t\cdot z}$ and $w^-_{k+1}(t,z)=P^-_{k+1}(t,-z) z^{-k-1}e^{-t\cdot z}$ be the (adjoint) wave function, corresponding to $\tau_k$ and $\tau_{k+1}$, i.e., given by  (\ref{wwave}) and (\ref{wwwave}) satisfying  (\ref{modKP3}) for $\ell=k+1$.
Then 
\begin{equation}
\label{**2}
P_{k+1}^+(t,\partial)\circ \partial= {\phi_k(t)}\partial \circ \frac1{\phi_k}(t) P^+_{k}(t,\partial)\, 
\end{equation}
and
\begin{equation}
\label{eqp55}
L_{k+1}=\phi_k(t)\partial\circ\frac1{\phi_k(t)} L_k \circ \phi_k(t)\partial^{-1}\circ\frac1{\phi_k(t)}.
\end{equation}
\end{proposition}
{\bf Proof.} If we divide equation (\ref{mod1}) by $\tau_k(t)\tau_{k+1}(y)$, we obtain
\begin{equation}
\label{**1}
{\rm Res}\,   w^+_k(t,z)\phi_k(y)w^-_{k+1}(y,z) dz=\phi_k(t)\frac1{\phi_k(y)}\, .
\end{equation}
which is equivalent to (\ref{modKP3}). Using Lemma \ref{L1} and (\ref{PP-}), we deduce
that 
\[
P^+_{k}(t,\partial)\circ \partial^{-1}\circ P^+_{k+1}(t,\partial)^{-1}=\phi_k(t)\partial^{-1}\circ \frac1{\phi_k(t)},
\]
which gives (\ref{**2}).
Then (\ref{eqp55}) follows from (\ref{Laxop}).\hfill$\square$
\\
\
 \\
The converse also holds:
\begin{proposition}
\label{Pkkk}
Let $\phi^+(t)$ be an eigenfunction and $\phi^-(t)$ be an adjoint eigenfunction for the Lax operator $L=P\partial P^{-1}$, i.e. $L$ satisfies (\ref{Lax}), where $P$ is a dressing operator, satisfying the Sato-Wilson equation (\ref{Sato}), then 
\[
Q=\phi^+(t)\partial \circ \frac1{\phi^+(t)}P\ \mbox{ and }R=\frac1{\phi^-(t)}\partial^{-1} \circ \phi^-(t)P
\]
also satisfy (\ref{Sato})
and both
\[
Q\circ \partial \circ Q^{-1}\ \mbox{ and }R \circ \partial\circ  R^{-1}
\]
are Lax operators.
\end{proposition}
For a proof of this proposition, see pages 499 and 500 of \cite{HL}.\\
\ \\
{\bf Proof of Proposition \ref{Matv}.}
We will only consider the case of eigenfunctions. The proof for adjoint eigenfunctions is similar.
Use the previous Proposition, then 
\[
{\rm Res}\, Qe^{t\cdot z}(P^{*})^{-1}e^{y\cdot z}dz
=\phi^+(t)\partial_{t_1} \circ \frac1{\phi^+(t)}{\rm Res}\, Pe^{t\cdot z}(P^{*})^{-1}e^{-y\cdot z}dz=0\, .
\]
Hence the wave function $Qe^{t\cdot z}$ and the adjoint wave function $(P^{*})^{-1}e^{-y\cdot z}$ satisfy the 1-st modified KP hierarchy, (\ref{modKP3}) for $k=\ell+1$.
Therefore, $Pe^{t\cdot z}$ and $(Q^{*})^{-1}e^{-y\cdot z}$ satisfy (\ref{mod2}), i.e.,
\[
{\rm Res}\, Pe^{t\cdot z}Q^{*-1}e^{y\cdot z}dz
=\frac{\phi^+(t)}{\phi^+(y)}\, .
\]
Let $\tau$ be the tau-function which corresponds to $P$ and $\tau_1$ be the tau-function that corresponds to $Q$, then 
\[
{\rm Res}\,   z^{-1}\tau(t-[z^{-1}])\tau_{1}(y+[z^{-1}]) \exp\left(\sum_{i=1}^\infty (t_i-y_i)z^i\right)dz=\tau(t)\phi^+(t) \frac{\tau_1(y)}{\phi^+(y)}\, ,
\]
which must be equation (\ref{mod1}).
Thus 
$\tau_1(t)=\phi^+(t) \tau(t)$.\hfill $\square$
\\
\
\\ \indent
Define 
\[
\phi^+_k(t)=\phi_k(t)\ \left(\mbox{resp. } \phi_k^-(t)=\frac1{\phi_{-k-1}}\right)\  \mbox{for } k\ge 0\, ,
\]
which are eigenfunctions for $L_k$ (resp. adjoint eigenfunctions for $L_{-k}$).
 Then by Proposition \ref{prop6},
\begin{equation}
\label{tauphi}
\phi_k^+(t)=\frac1{\phi_{-k-1}^-(t)}=\frac{\tau_{k+1}(t)}{\tau_k(t)},
\end{equation}
and (by (\ref{Pll}) and  Proposition \ref{prop6})
\begin{equation}
\label{LLL2}
\partial+v_k(t)=
\begin{cases}
\partial -\partial (\log \phi_k^+(t))
=\phi_k^+(t)\partial\circ\frac1{\phi_k^+(t)}& \mbox{for }k\ge 0,\\
\partial +\partial (\log \phi_{-k-1}^-(t))=\frac1{\phi_{-k-1}^-(t)}\partial\circ\phi_{-k-1}^-(t)&  \mbox{for }k< 0,\\
\end{cases}
\end{equation}
and
\begin{equation}
\label{11}
\begin{aligned} 
  w^{\pm}_ {k+1}(t,z)&=\pm (\phi_k^+(t)^{\pm 1}\partial^{\pm 1}\circ\phi_k^+(t)^{\mp1})
   w^\pm_k(t,z),\\
   w_{-k-1}^\pm(t,z)&=\pm (\phi_k^-(t)^{\mp 1}\partial^{\mp 1}\circ\phi_k^-(t)^{\pm 1})
    w^\pm_{-k}(t,z).
\end{aligned}
\end{equation}
It is clear that the first and the second formulation of MKP imply\\ 
\ 
\\
{\bf The third formulation of the MKP hierarchy: }
\\
{\it Let $W=\mathbb{C}[u_i^{(n)}, {\phi_j^\pm}^{(n)}|\,  i\in\mathbb{Z}_{\ge 1},\ j, n\in\mathbb{Z}_{\ge 0}]$ be the algebra of differential polynomials in $u_i$
 and $\phi^\pm_j$, where $\partial u_i^{(n)}=u_i^{(n+1)}$, $\partial {\phi_j^\pm}^ {(n)}={\phi_j^\pm}^ {(n+1)}$.
Let $L_0(\partial )=\partial +u_1(t)\partial^{-1} +u_2(t)\partial^{-2} \ldots \in W((\partial^{-1}))$ be a pseudo-differential operator. 
Then  the MKP hierarchy is the following system of evolution equations in $W$:
\begin{equation}
\label{third}
\frac{\partial L_0(\partial)}{\partial t_j} =[(L_0(\partial)^j)_+,L_0(\partial)], \quad \frac{\partial \phi^+_i}{\partial t_j}=\left(L_{i}(\partial)^j\right)_+(\phi_i^+),\quad
\frac{\partial \phi^-_i}{\partial t_j}=-\left(L_{-i}(\partial)^{j}\right)^*_+(\phi_i^-)
\end{equation}
for $j\in\mathbb{Z}_{\ge 1}$ and  $ i\in\mathbb{Z}_{\ge 0}$,
where  the $L_i$ and $L_{-i}$, for $i>0$, are defined by
\[
L_i=\phi_{i-1}^+ \partial\circ \frac1{\phi_{i-1}^+} L_{i-1}\circ \phi_{i-1}^+\partial^{-1}\circ  \frac1{\phi_{i-1}^+},\quad
L_{-i}=\frac1{\phi_{i-1}^-}\partial^{-1}\circ \phi_{i-1}^- L_{1-i}\circ\frac1{\phi_{i-1}^-}\partial \circ\phi_{i-1}^-\, .
\]
} 
\begin{theorem}
\label{theoremnew}
All three formulations of the MKP are equivalent.
\end{theorem}
{\bf Proof}
Assume the third formulation of MKP holds. Define  for $i\ge 0$ the function  $v_i=-\partial \log\phi_i^+$ and $v_{-i-1}=\partial \log\phi_i^-$.
Then 
\[
w^+_{i+1}(t,z)=\phi_i^+(t)\partial\circ\frac1{\phi_i^+(t)}(w^+_{i}(t,z))=(\partial+v_i(t))(w^+_{i}(t,z))
\]
is a wave function for $L_{i+1}=(\partial+v_i(t))L_i (\partial+v_i(t))^{-1}$.
One finds  similar wave functions and relations between these wave functions if $i<0$.
Hence,
the same proof as the proof of Theorem \ref{theorem2} gives the second equation of (\ref{Dickey}). Equation (\ref{LLL}) is obvious.\hfill$\square$\\
\ 
\\
\indent 
Now, for $i>0$,  the tau-function is equal to (by (\ref{tauphi}))
\begin{equation}
\label{taui}
\tau_{\pm i}=\phi_{i-1}^\pm\tau_{\pm (i-1)}=\phi_{i-1}^\pm\phi_{i-2}^\pm\tau_{\pm (i-2)} =\cdots =\phi_{i-1}^\pm\phi_{i-2}^\pm\cdots \phi_0^\pm\tau_0\, ,
\end{equation}
and  the (adjoint) wave function $w_{\pm i}^\pm(t,z)=M_{\pm i}(t, \partial) \left(w_0^\pm (t,z)\right)$, where $M_0=1$  and by  (\ref{11}) and (\ref{LLL2}):
\begin{equation}
\label{Mi}
\begin{aligned}
M_{\pm i}(t, \partial)
&=(\pm\partial +v_{\pm (i-\frac12 \mp\frac12)})\circ M_{\pm (i-1)}(t, \partial)\\
&=\pm \phi_{i-1}^\pm \partial\circ
\frac{1}{\phi_{i-1}^\pm} M_{\pm (i-1)}(t, \partial)\\
&=\phi_{i-1}^\pm \partial\circ
\frac{1 }{\phi_{i-1}^\pm} \phi_{i-2}^\pm\partial\circ \frac1{\phi_{i-2}^\pm} M_{\pm (i-2)}(t, \partial)\\
&=\cdots \\
&=(\pm 1)^i\phi_{i-1}^\pm \partial\circ
\frac{\phi_{i-2}^\pm }{\phi_{i-1}^\pm}\partial \circ \frac{\phi_{i-3}^\pm}{\phi_{i-2}^\pm}
\partial\circ\cdots \circ \frac{\phi_{0}^\pm}{\phi_{1}^\pm}\partial \circ\frac1{\phi_{0}^\pm}\, ,
\end{aligned}
\end{equation}
is an $i$-th order differential operator. 
Using the connection between the wave function and adjoint wave function we have , $w_{- i}^+(t,z)=M_{- i}^{*-1}(t, \partial) \left(w_0^+ (t,z)\right)$
and using the relation between the wave function and the Lax operator (\ref{Laxop}), we find
\begin{equation}
\label{66}
L_{i}=M_{i}\circ L_{0}\circ M_{i}^{-1}\ \mbox{ and }L_{-i}=(M_{-i}^{*})^{-1}\circ L_{0}\circ M_{-i}^*\, .
\end{equation}

\
\\
\noindent
In the polynomial case, using the boson-fermion correspondence $\sigma$, it is not difficult to find these (adjoint) eigenfunctions. 
We know from the results of \cite{KP} that if $\sigma^{-1}(\tau_nq^n)=f_n\in {\cal O}_{n}$, then $\sigma^{-1}(\tau_{n+1}q^{n+1})=w\wedge f_n$ for some
$w=\sum_i a_i e_i\in \mathbb{C}^\infty$.
We have 
\[ 
f_{n+1}=w\wedge f_n=\left(\sum_i a_i e_i\right)\wedge f_n=\sum_ia_i \psi_{-i+\frac12}^+(f_n)={\rm Res}\, \sum_ia_i z^{-i}\psi^+(z) (f_n)dz \, ,
\]
since this holds for $f_n=|0\rangle$ and $ f_{n+1}=|n+1\rangle$.
Thus if we define $\phi^+_n(t)= {\rm Res}\, \sum_ia_i z^{-i} w^+_{n}(t,z) dz$, then by (\ref{wave}) we find that
\begin{equation}
\label{taun+1}
\begin{aligned}
\tau_{n+1}q^{n+1}=&\sigma\left({\rm Res}\, \sum_ia_i z^{-i}\psi^+(z) (f_n)dz \right)\\
=&{\rm Res}\, \sum_ia_i z^{-i} \sigma\psi^+(z) \sigma^{-1}dz\,  \tau_n q^n \\
=&{\rm Res}\, \sum_ia_i z^{-i} w^+_{n}(t,z) dz\, \tau_n q^{n+1}\\
=&\phi^+_n\tau_n q^{n+1}\, ,
\end{aligned}
\end{equation}
hence
\begin{equation}
\label{ttau}
\tau_{n+1}=\phi^+_n\tau_n,\ \mbox{where }
        \phi^+_n={\rm Res}\,\sum_ia_i z^{-i} w^+_{n}(t,z) dz\, .
\end{equation}
Since $f_{n-1}=\sum_ib_i\psi_{i+\frac12}^- (f_n)$, with $b_i\in\mathbb{C}$,  in a similar way we find
\begin{equation}\label{tttau}
\tau_{-n-1}=\phi^-_n \tau_{-n},\ \mbox{where }\phi^-_n(t)= {\rm Res}\, \sum_i b_i z^{i} w^-_{-n}(t,z) dz\, .
\end{equation}
Thus we have the following
\begin{lemma}
\label{lemmaeigen}
In the polynomial setting every (adjoint) eigenfunction is of the form (\ref{qwave}).
\end{lemma}

Observe that since $\phi_1^\pm={\rm Res}\, f(z) w_{\pm 1}^\pm(z) dz$ for some $f(z)$, we find that if we define $q_0^\pm=\phi_0^\pm$ and $q_1^\pm={\rm Res}\, f(z) w_0^\pm(z) dz$, which are both  (adjoint) eigenfunctions of $L_0$, then using (\ref{11}) we deduce that
\[
\begin{aligned}
\phi_1^\pm=& {\rm Res}\, f(z) w_{\pm 1}^\pm(z) dz\\
=& \pm{\rm Res}\, f(z)\phi_0^\pm\partial  \left(\frac{w_0^\pm(z)}{\phi_0^+}\right) dz
\\
=&\pm q_0^\pm\partial \left( \frac{q_1^\pm}{q_0^\pm}\right)\\
=&\pm \left(\partial(q^\pm _1)- \frac{q_1^\pm}{q_0^\pm}\partial(q_0^\pm )\right)\, .
\end{aligned}
\]
Thus 
\[
\tau_{\pm 2}=\phi^\pm_0\phi^\pm_1\tau_0=\pm q_0^\pm\left(\partial(q^\pm _1)- \frac{q_1^\pm}{q_0^\pm}\partial(q_0^\pm )\right)\tau_0=
\pm\det
\begin{pmatrix}
q_0^\pm &q_1^\pm\\
\partial(q_0^\pm)&\partial(q_1^\pm)
\end{pmatrix}
\tau_0\, .
\]
Note that we can remove the possible minus sign in front of the determinant. If $\tau_2$ is a tau-function, then a multiple of $\tau_2$ is also a tau-function. From now on we will always do so, i.e. forget about the sign of the tau-function.

Using formula (\ref{Mi}), we deduce that
\[
M_{\pm 1}=\pm \phi_0^\pm\partial \circ \frac{1}{\phi_{0}^\pm}
\]
and
\[
\begin{aligned}
M_{\pm 2}=&\phi_1^\pm\partial \circ \frac{\phi_{0}^\pm}{\phi_{1}^\pm}\partial \circ\frac1{\phi_{0}^\pm}\\
=&\frac1{q_0^\pm}\left(q_0^\pm\partial(q^\pm_1)-q_1^\pm\partial(q_0^\pm)\right)\partial \circ \frac{(q_{0}^\pm)^2}{q_0^\pm\partial(q^\pm_1)-q_1^+\partial(q_0^\pm)}\partial \circ\frac1{q_{0}^\pm}\\
=&\left(\det
\begin{pmatrix}
q_0^\pm&q_1^\pm\\
\partial(q_0^\pm)&\partial(q_1^\pm)
\end{pmatrix}\right)^{-1}\det
\begin{pmatrix}
q_0^\pm&q_1^\pm&1\\
\partial(q_0^\pm)&\partial(q_1^\pm)&\partial\\
\partial^2(q_0^\pm)&\partial^2(q_1^\pm)&\partial^2
\end{pmatrix}\, .
\end{aligned}
\]
Continuing in this way, see e.g.  Theorem 5.1  of \cite{HL} for more details,  it is possible to express $M_{\pm i}$ in terms of certain (adjoint) eigenfunctions $q^\pm_k(t)$ of the operator $L_0$, i.e. 
if 
\[
\phi_k^\pm ={\rm Res} f_k(z)w_{\pm k}^\pm dz\, ,
\]
for some $f_k(z)\in\mathbb{C}[z,z^{-1}]$, then we define
\[
q_k^\pm ={\rm Res} f_k(z)w_0^\pm dz\, .
\]
These $q^\pm_k(t)$ are (adjoint) eigenfumctions for $L_0(\partial)$ by (\ref{qwave}). 
We have the following formulas:
\begin{equation}
\label{tauW}
\tau_{\pm i}=
W_{\pm i}\tau_0\ \mbox{and }w_{\pm i}^\pm=M_{\pm  i}(w_0^\pm)\ \mbox{and }
w^+_{-i}=(M_{-i}^{*})^{-1}(w_0^+),
\end{equation}
where  $M_{\pm i}= (\pm 1)^iW_{\pm i}(\partial)/W_{\pm i}$, and
\begin{equation}\label{Wronskian}
W_{\pm i}(\partial)= \det
\begin{pmatrix}
q_0^\pm&\cdots&q_{i-1}^\pm&1\\
\partial(q_0^\pm)&\cdots&\partial(q_{i-1}^\pm)&\partial\\
\vdots&\ddots&\vdots&\vdots\\
\partial^{i}(q_0^\pm)&\cdots&\partial^{i}(q_{i-1}^\pm)&\partial^{i}
\end{pmatrix}
\ \text{and }
W_{\pm i}= \det
\begin{pmatrix}
q_0^\pm&\cdots&q_{i-1}^\pm\\
\partial(q_0^\pm)&\cdots&\partial(q_{i-1}^\pm)\\
\vdots&\ddots&\vdots\\
\partial^{i-1}(q_0^\pm)&\cdots&\partial^{i-1}(q_{i-1}^\pm)
\end{pmatrix}
\end{equation}
are Wronskian determinants. The determinants $W_{\pm i}(\partial)$ are computed by expanding along the last column, putting the cofactors to the left of the $\partial^j$'s.

Let us prove  the formulas of (\ref{tauW}).
If $\tau_{\pm i}=W_{\pm i}\tau_0$, then  
\[
\begin{aligned}
\tau_{\pm i\pm 1}&=\phi^\pm_i\tau_{\pm i}\\
&={\rm Res}\,  f_i(z)w^{\pm}_{\pm i}dz\, W_{\pm i}\tau_0\\
&={\rm Res}\,  f_i(z)M_{\pm i}(w^{\pm}_0)dz\, W_{\pm i}\tau_0\\
&={\rm Res}\,  f_i(z)W_{\pm i}(\partial)(w^{\pm}_{0})dz\, \tau_0\\
&=W_{\pm i}(\partial)({\rm Res}\,  f_i(z)w^{\pm}_{0}dz) \tau_0\\
&=W_{\pm i}(\partial)(q^\pm_{i}) \tau_0\\
&=W_{\pm (i+ 1)}\tau_0\, .
\end{aligned}
\]
Thus  
\[
\phi_i^\pm=\frac{W_{\pm (i+1)}}{W_{\pm i}},
\] and using this,
we find that 
\[
\begin{aligned}
		w^\pm_{\pm (i+ 1)}&=\pm \phi_{ i}^\pm \partial \circ \frac{1}{ \phi_{ i}^\pm}(w^\pm_{\pm i})\\
&=(\pm 1)^{i+1} \frac{W_{\pm (i+ 1)}}{W_{\pm i}}\partial \circ \frac {W_{\pm i}}{W_{\pm (i+ 1)}}\circ M_{\pm i}(w^\pm_0)\\
&=(\pm 1)^{i+1} \frac{W_{\pm (i+ 1)}}{W_{\pm i}}\partial\circ   \frac {W_{\pm i}}{W_{\pm (i+ 1)}}\left(\frac {W_{\pm i}(\partial)(w^\pm_0)}{W_{\pm i}}\right)\\
                &
                =(\pm 1)^{i+1}\frac{W_{\pm (i+1)}(\partial)(w^\pm_0)}{W_{\pm (i+1)}}\\
&=M_{\pm (i+ 1)}(w^\pm_0).
\end{aligned}
\]
The next to the last equality follows from  Crum's Identity  for Wronskian determinants (which is in fact the Desnanot-Jacobi identity for Wronskians, see \cite{C}, section 3):
\begin{equation}
\label{Crum}
W_{\pm (i+1)}\partial\circ W_{\pm i}(\partial)
-\partial (W_{\pm (i+ 1)})W_{\pm i}(\partial)=W_{\pm i}W_{\pm (i+ 1)}(\partial)\, .
\end{equation}
Thus $w_{-i}^+= (M_{-i}^*)^{-1} (w_0^+)$.
Now by (\ref{66}) we find that
\begin{equation}
\label{LM}
\begin{aligned}
 L_{i}&=M_{i}\circ L_{0}\circ M_{i}^{-1}=W_{ i}(\partial)/W_{ i}\circ L_{0}\circ\left(W_{ i}(\partial)/W_{ i}\right)^{-1}\\
L_{-i}&=M_{-i}^{*-1}\circ L_{0}\circ M_{-i}^*=\left(W_{- i}(\partial)/W_{- i}\right)^{*-1}\circ L_{0}\circ \left(W_{- i}(\partial)/W_{- i}\right)^{*}.
\end{aligned}
\end{equation}
\begin{remark}
\label{R}
 Let $i\geq 0$ and let $f_i=\sigma^{-1}(\tau_i(t)q^i)$.
Then  $f_i\in {\cal O}_i$,
which means that  
\begin{equation}
\label{fi}
\begin{aligned}
  f_i
=v_i\wedge v_{i-1}\wedge\cdots\wedge v_2\wedge v_1\wedge f_0,\   \mbox{where }
 v_j=\sum_s a_{sj}e_s,\, f_0\in {\cal O}_0,
\end{aligned}
\end{equation}
and the eigenfunctions of $L_{j}$ are of the form
\[
\phi_{j}^+(t)
 = {\rm Res}\, w_{j}^+(t,z)\sum_i a_{i,j+1}z^{-i}dz.
\]
Hence, this eigenfunction is determined by $ w_{j}^+(t,z)$ and by $v_{j+1}$.
Define 
\[
q_{j}^+(t)
 = {\rm Res}\, w_{0}^+(t,z)\sum_i a_{i,j+1}z^{-i}dz.
\] 
%
Since $M_i$ is of the form (\ref{Mi}), $ \phi_0^+(t)=q_0^+(t)$ is in the kernel of $M_i$. However, if we reorder the $v_j$'s in (\ref{fi}) we get the same element up to a sign.
This gives different eigenfunctions $\phi_j^+$  and different $L_j$ for $j=1,2,\ldots, i-1$, but $M_i$ is the same and $L_i$ is the same. Hence we can put every $v_j$ in (\ref{fi}) just before $f_0$,
which means that the new $f_1=v_j\wedge f_0$, thus we get a new eigenfunction $\phi_0^+$ which is now equal to $q^+_j(t)$. Moreover, if $f_i\not =0$, then
$q^+_j(t)\not = 0$.
%
Thus $q_0^+(t),q_1^+(t),\ldots, q_{i-1}^+(t)$ are  nonzero eigenfunctions for $L_0$ which are all in the  kernel of $M_i$, and clearly must be linearly independent otherwise the element $f_i$ would be 0. Similarly 
\[
f_{-i}=v_{-i}(v_{1-i}(\cdots (v_{-2}(v_{-1}(f_0)\cdots )\, ,
\]
where  $v_j=\sum_i b_{ij}\psi_{i+\frac12}^-$.
Then
\[ 
\phi_{j-1}^-(t)=  {\rm Res}\, w_{1-j}^-(t,z)\sum_i b_{i,-j}z^{i}dz,\ \mbox{and}\quad
q_{j-1}^-(t)=  {\rm Res}\, w_0^-(t,z)\sum_i b_{i,-j}z^{i}dz\,, 
\]
and all $q_j^-(t)$ for $0\le j<i$ are in the kernel of $M_{-i}$.
\end{remark}
{\bf The fourth  formulation of the MKP hierarchy: }
\\
{\it Let $V=\mathbb{C}[u_i^{(n)}, {q_j^\pm}^{(n)}| i\in\mathbb{Z}_{\ge 1},\ j, n\in\mathbb{Z}_{\ge 0}]$ be the algebra of differential polynomials in $u_i$ and $q^\pm_j$. 
Let $L_0=\partial +u_1(t)\partial^{-1} +\ldots \in V((\partial^{-1}))$ be a pseudo-differential operator. Then  the MKP hierarchy is the following system of evolution equations in $V$:
\begin{equation}
\label{fourth}
\frac{\partial L_0(\partial)}{\partial t_j} =[(L_0(\partial)^j)_+,L_0(\partial)], \quad \frac{\partial q^+_i}{\partial t_j}=\left(L_{0}(\partial)^j\right)_+(q_i^+),\quad
\frac{\partial q^-_i}{\partial t_j}=-\left(L_{0}(\partial)^{j}\right)^*_+(q_i^-)\, .
\end{equation}
}
Now we are able to prove the following 
\begin{theorem}\label{theorem3}
In the polynomial setting, all four formulations of MKP are equivalent.
\end{theorem}

\noindent
{\bf Proof.} 
It suffices to establish the equivalence between the third and fourth formulation.
To obtain the fourth formulation from the third, we use the fact that 
if $\phi_i^\pm(t)$ is given, then by Lemma \ref{lemmaeigen} this   (adjoint) eigenfunction for $L_{\pm i}$ is equal to
\[
\phi_i^\pm(t)={\rm Res}\, f^\pm (z) w_{\pm i}^\pm(z) dz\ \mbox{ for some }f(z)\in \mathbb{C}((z^{-1}))\, .
\]
Then we define the $q_i^\pm (t)$ of the fourth formulation by
\[
q_i^\pm (t)={\rm Res}\, f^\pm (z) w_0^\pm(z) dz\ \mbox{ for the same }f(z)\in \mathbb{C}((z^{-1}))\, ,
\]
which now is an (adjoint) eigenfunction for $L_0$. This $q^\pm_i(t)$ for $i\ge 0$ is (by Remark \ref{R}) in the kernel of $M_{\pm j}$ (defined in (\ref{Mi}))
for $j\ge i$,
and since it is an (adjoint) eigenfunction for $L_0$, it satisfies the second (third) formula of (\ref{fourth}).
Hence this establishes  the fourth formulation of MKP.

Assume the fourth formulation holds. 
Define $\phi^\pm_{\pm n}=(-1)^n\frac{W_{\pm n\pm 1}}{W_{\pm n}}$; together with $L_0$ they form the data of the third formulation.
Since $q^\pm_i$ is an (adjoint) eigenfunction of $L_0$, then by Lemma \ref{lemmaeigen} there exist functions $f^\pm_i(z)\in\mathbb{C}((z^{-1}))$, such that
\begin{equation}
\label{help}
q^\pm_i(t)={\rm Res}\, f^\pm_i(z)w_0^\pm(t,z)dz\, .
\end{equation}
Let $\tau_0$ be the tau-function for $L_0$. Since $q^\pm_0=\phi_0^\pm$ is an (adjoint) eigenfunction of $L_0$,  by Proposition \ref{Matv}, the tau-functions for $L_{\pm}$ are
\[
\tau_{\pm 1}=W_{\pm 1}\tau_0=\phi^{\pm }_0\tau_0\, .
\] 
The corresponding 
 (adjoint) wave functions are (by Propositions  \ref{Pkk} and \ref{Pkkk})
\begin{equation}
\label{H1}
\begin{aligned}
w_1^+(t,z)&=M_1(w_0^+(t,z))=\phi_0^+(t)\partial \circ\frac1{\phi_0^+(t)}(w_0^+(t,z))\, ,\\
w_{-1}^-(t,z)&=M_{-1}(w_0^-(t,z))=-{\phi_0^-(t)}\partial \circ\frac1{\phi_0^+(t)}(w_0^-(t,z))\, ,
\end{aligned}
\end{equation}
where $M_{\pm 1}$ is given by (\ref{tauW}).
The corresponding Lax operator $L_{\pm 1}$ is defined by (\ref{LM}), which is the same as $L_{\pm 1}$  in the third formulation, because of (\ref{H1}).
Let
\[
\begin{aligned}
\phi^\pm_{ 1}(t)
&={\rm Res}\, f^\pm_1(z)w_{\pm 1}^\pm(t,z)dz\\
&=\pm{\rm Res}\, f^\pm_1(z)\frac{W_{\pm 1} (\partial)(w^\pm_0(t,z))}{W_{\pm 1}}\\
&=\pm\frac{W_{\pm 1} (\partial)(q^\pm_1(t))}{W_{\pm 1}}\\
&=\pm\frac{W_{\pm 2}}{W_{\pm 1}}
\, ,
\end{aligned}
\]
where $ f^\pm_1(z)$ is given by (\ref{help}), which are non-zero by Remark \ref{R}.
Now, $\phi^\pm_{ 1}(t)$ is an (adjoint) eigenfunction for $L_{\pm 1}$, hence the second equation of (\ref{third}) holds for $L_{\pm 1}$ and $\phi_{1}^\pm$.
Thus (by (\ref{tauW}))  we obtain that the  
tau-functions for $L_{\pm 2}$ are equal to 
\[
\tau_{\pm 2}=W_{\pm 2}\tau_0=\frac{W_{\pm 2}}{W_{\pm 1}}W_{\pm 1}\tau_0=\pm\phi_1^\pm  \tau_{\pm 1}\, .
\]
The corresponding
(adjoint) wave functions are given by (\ref{tauW})  and (\ref{Wronskian}), and
we have
\[
w_2^+(t,z)=M_2(w_0^+(t,z))=\frac{W_2(\partial)(w_0^+(t,z))}{W_2
}\, . \]
By Crum's identity (\ref{Crum}) we find that
\begin{equation}
\label{H2}
\begin{aligned}
w_2^+(t,z)
=&\frac{W_2}{W_1}\partial\circ
\frac{W_1}{W_2}\left(\frac{W_1(\partial)(w_0^+(t,z)}{W_1}\right)\\
&=\phi_1^+(t)\partial \circ\frac1{\phi_1^+(t)}\circ M_1(w_0^+(t,z))\\
&=\phi_1^+(t)\partial \circ\frac1{\phi_1^+(t)}(w_1^+(t,z)),\\
w_{-2}^-(t,z)&=M_{-2}(w_0^-(t,z))=-{\phi_1^-(t))}\partial \circ\frac1{\phi_1^-(t)}(w_{-1}^-(t,z))\, .
\end{aligned}
\end{equation}
The corresponding Lax operator $L_{\pm 2}$ is defined by (\ref{LM}), which is the same as the one  in the third formulation, because of (\ref{H2}).
Let
\[
\phi_{ 2}^\pm(t)={\rm Res}\, f^\pm_2(z)w_{\pm 2}^\pm(t,z)dz
={\rm Res}\, f^\pm_2(z)\frac{W_{\pm 2} (\partial)(w^\pm_0(t,z))}{W_{\pm 2}}dz
=\frac{W_{\pm 3}}{W_{\pm 2}}
\, ,
\]
where again $ f^\pm_2(z)$ is given by (\ref{help}). This is again an (adjoint) eigenfunction for $L_{\pm 2}$ and hence the second equation of (\ref{third}) holds for $L_{\pm 2}$ and $\phi_{2}^\pm$.
Continuing along these lines gives
the third formulation and hence we have proved that all four formulations are equivalent.
\hfill $\square$
\\

\section{Polynomial solutions of MKP}
\label{S7}
We are now going to  construct polynomial tau-functions for MKP.
We assume that $f_0=|0\rangle $ which means that $\tau_0(t)=1$, $w^\pm (t,z)=e^{\pm t\cdot z}$ and $L_0=\partial$. We construct a $L_0=\partial$ eigenfunction by the procedure described in  Example \ref{EX} at the beginning of  Section \ref{Seigen}. 
Since $f_1=w\wedge f_0=w\wedge |0\rangle$ and the vacuum is given by (\ref{vac}), such a $w$  can be chosen of the form
$w=\sum_{j=0}^\infty a_j e_{j+1}$, thus the corresponding eigenfunction $q^+(t)=\tau_1(t)$ is of the form (see Example \ref{EX})
\[
q^+(t)={\rm Res }\sum_{j=0}^\infty a_j z^{-j-1}e^{ t\cdot z}dz\, .
\]
A similar construction is possible for the adjoint eigenfunction, in fact we have that
all (adjoint) eigenfunctions are of the form 
\begin{equation}
\label{1111}
q^\pm_i(t)={\rm Res}\, f_i^\pm(z) e^{\pm t\cdot z}dz,\ \mbox{for some  }  f_i^\pm(z)=\sum_{j=0}^\infty a_{ji}^\pm z^{-j-1} \, .
\end{equation}
Since  $\tau_0=1$ and $\tau_n=W_n\tau_0$ (see  (\ref{tauW})), the  corresponding tau-function is equal to $\tau_n=W_n$, for $n\in \mathbb{Z}$,  the Wronskian determinant of the (adjoint) eigenfunctions.
Now using the  elementary Schur polynomials, which are defined by
\begin{equation} 
\label{elSchur}e^{t\cdot z}=\sum_{j=0}^\infty s_j(t)z^j\, , \end{equation}
we find (see (\ref{1111})) that 
\[
q^\pm_i(t)={\rm Res}\, f_i^\pm(z) e^{\pm t\cdot z}dz=\sum_{j=0}^\infty a_{ji} ^\pm
s_j(\pm t)\, .
\]
One obtains polynomial tau-functions by taking 
$f_i^\pm(z) =\sum_{j=0}^{M^\pm _i} a^\pm_{ji}z^{-j-1}$. To simplify notation we shall sometimes drop the superscrips $\pm$. Without loss of generality we  may assume that $a_{M_i ,i}=1$, then 
\[
q^\pm_{i}(t)=s_{M_i}(\pm t)+ 
\sum_{j=0}^{M _i-1} a_{ji}s_j(\pm t).
\]
One can find recursively constants $c_i =(c_{1i},c_{2i}, \dots, c_{M_i i})$, such that 
\begin{equation}
\label{QQ}
q^\pm_{i}(t)=s_{M_i}(\pm t)+\sum_{j=0}^{M_i -1} a_{ji }s_j(\pm t)=
s_{M_i}(\pm (t+ c _{i}))\, .
\end{equation}
Indeed, since,
$
s_{M_i}(t+c_i)=\sum_{j=0}^{M_i} s_j(c_i)s_{M_i-j}(t)\, ,
$
which follows immediately from (\ref{elSchur}), one has to solve equations of the form
$
s_j(c_i)=a_{M_i-j,i}
$ and this can be done recursively since $s_j(c_i)=c_{ji}+p_j(c_{1i},\ldots, c_{j-1,i})$, where $p_j$ is some polynomial. First, determine $c_{1,i}$, which is determined by $a_{M_i-1,i}$,  then $c_{2,i}$, which is determined by $a_{M_i-2,i}$ and $c_{1i}$, then  $c_{3i}$, which is determined by $a_{M_i-3,i}$, $c_{1i}$ and $c_{2i}$, etc. 
In fact there is an explicit formula for these constants. 
Since 
\[
1+\sum_{j=1}^{M_i}a_{M_i-j,i}z^j=\sum_{j=0}^M s_j(c_i)z^j\, ,
\]
which is equal to the first  $M_i+1$ terms of $\exp(\sum_{j=1}^{M_i} c_{ji} z^j)$, the logarithm of this gives that
\[
\begin{aligned}
&\sum_{\ell=1}^{M_i} c_{\ell i} z^\ell i +\mbox{ higher order terms}=\log\left(1+\sum_{k=1}^{M_i} a_{M_i-k,i}z^k\right).\\
\end{aligned}
\]
Hence
\[
c_{ki}=-\sum_{m_1+2m_2+\cdots +km_k=k \atop m_1\ge 0,m_2\ge 0,\ldots, m_k\ge 0}\prod_{j=1}^k \frac{(-a_{M_i-j,i})^{m_j}}{m_j}\, .
\]

Since $\tau_0=1$ and $\tau_{\pm n}=W_{\pm n}\tau_0$, we have (see (\ref{Wronskian}) and (\ref{QQ}))
that
\begin{equation}
\label{eqP6}
\begin{aligned}
 \tau_{\pm n}(t)=&
W(q_0^\pm(t), q_1^\pm (t),\ldots, q_{n-1}^\pm (t))\\
=&W\left(s_{M^\pm_0}(\pm t+c^\pm_{0}), s_{M_1^\pm}(\pm t+c^\pm_{1}),\ldots ,s_{M^\pm_{n-1}}(\pm t+c^\pm_{n-1})\right) ,
\end{aligned}
\end{equation}
where $W(\ )$ stands for the Wronskian determinant of those (adjoint) eigenfunctions, satisfies the KP hierarchy.
This shows that every function  of the form ({\ref{eqP6}) is a polynomial tau-function. Moreover, one has  the following remarkable
\begin{theorem}
\label{T7}
(a) All polynomial tau-functions of the KP hierarchy are, up to a constant factor, of the form
\begin{equation}
\label{Pt}
\tau_{\lambda_1,\lambda_2,\ldots,\lambda_k}(t;c_1,c_2,\ldots ,c_k)=\det \left( s_{\lambda_i+j-i}(t_1+c_{1,i},t_2+c_{2i},t_3+c_{3i},\ldots)\right)_{1\le i,j\le k}\, ,
\end{equation}
where $\lambda=(\lambda_1,\lambda_2, \ldots,\lambda_k)$ is a partition and
$c_i=(c_{1i},c_{2i},\ldots) \in\mathbb{C}^k$ are arbitrary.\\
(b) All polynomial tau-functions of the MKP hierarchy are the sequences
$(...,\tau_n, \tau_{n+1},...)$,  where each $\tau_n$ is, up to a constant factor, of the form   (\ref{Pt})), and $\tau_{n+1}$ is obtained from $\tau_n$, up to a constant factor, in one of the following three possible ways:
\begin{itemize}
\item $\tau_{\mu ,\lambda_1,\lambda_2,\ldots,\lambda_k}(t;d, c_1,c_2,\ldots ,c_k)$,
  with $\mu\geq \lambda_1$; 
\item $\tau_{\lambda_1-1,\lambda_2-1,\ldots,\lambda_i-1, \mu, \lambda_{i+1},\ldots \lambda_k}(t; c_1,c_2,\ldots ,c_i,d, c_{i+1},\ldots c_k)$, for $i=1,2,\ldots k$, with
  $\lambda_i>\mu\geq \lambda_{i+1}$;
\item $\tau_{\lambda_1-1,\lambda_2-1,\ldots,\lambda_k-1}(t;c_1,c_2,\ldots ,c_k)$.
\end{itemize}
Here  $d=(d_1,d_2,\ldots)$ is a set of constants connected to the part $\mu$ of the  partition, that appears in $\tau_{n+1}$, in the first two cases. In the third case one  has to delete $\lambda_j-1$'s
and the corresponding $c_j$'s, whenever $\lambda_j-1$ is equal to $0$.
\end{theorem}
{\bf Proof. (a)}  First reorder the functions in (\ref{eqP6}) such that $M_0>M_1>M_2>\cdots>M_{k-1}$, which leaves the tau-function unchanged up 
to a sign. If one writes out (\ref{eqP6}), (cf. (\ref{Wronskian})), where $q_i^+$ is an elementary Schur function $s_{M_i}$, using that $\frac{\partial^\ell s_{M_i} }{\partial t_1^\ell}=s_{M_i-\ell}$,  it is immediate to check
that the the Wronskian matrix of  (\ref{eqP6}) is the transposed of the matrix in:
\begin{equation}
\label{eqT7}
\tau_k(t)=\det \left( s_{M_{i-1}+j-k}(t_1+c_{1,i},t_2+c_{2i},t_3+c_{3i},\ldots)\right)_{ij}\, .
\end{equation}
Now, $\tau_n(t)$  is the image under  the map $\sigma$ in $B$  of the following element of $F^{(0)}$ (cf. (\ref{QQ}), where we remove the 
upper index +, to simplify notation): 
\[
\begin{aligned}
&
\left(e_{M_0+1-k}+\sum_{j=1}^{M_0 }a_{j-1,0}e_{j-k}\right)\wedge \cdots \wedge\left(e_{M_{k-1}+1-k}+\sum_{j=1}^{M_{k-1} }a_{j-1,k-1}e_{j-k}\right)\wedge e_{-k}\wedge e_{-k-1}\wedge\cdots
\\
&=R\left(
I+\sum_{\ell=0}^{k-1}\sum_{j=1}^{M _{\ell} } a_{j-1,\ell}  E_{j-k ,M_\ell+1-k}
\right)(e_{M_0+1-k}\wedge \cdots\wedge e_{M_{k-1}+1-k}\wedge e_{-k}\wedge e_{-k-1}\wedge  \cdots)\, .
\end{aligned}
\]
Recall that (see \cite{KP})  
\[
\sigma(e_{M_0+1-k}\wedge e_{M_1+1-k}\wedge\cdots\wedge e_{M_{k-1}+1-k}\wedge e_{-k}\wedge e_{-k-1}\wedge e_{-k-2}\wedge \cdots))=s_{\lambda}(t)\,  ,
\]
where
\[s_\lambda(t)=\det(s_{\lambda_i +j-i}(t))_{1\leq i,j\leq k}\]
is the Schur polynomial, corresponding to the partition $\lambda=(\lambda_1,\lambda_2, \ldots,\lambda_k)$, with $\lambda_i=M_{i-1}+i-k$. Thus (\ref{eqT7}) lies in
$\sigma R(U)\sigma^{-1}\cdot s_\lambda(t)$, where $R$ is the representation of $GL_\infty$
in $F$ (see Section 2), so that $\sigma R \sigma^{-1}$ is the corresponding representation in $B$, and $U$ is the subgroup of $GL_\infty$, consisting of upper triangular
matrices with $1$'s on the diagonal.

We will next show that the dimension of the space of all polynomials of the form (\ref{eqT7}) is $-\frac12k(k-1)+\sum_{i=0}^{k-1} M_i$, or in terms 
of the corresponding partition $\lambda$, it is 
$|\lambda |=\lambda_1+\lambda_2+\cdots+\lambda_k$. To show this, we first calculate the degrees of freedom of such a solution. 
Since it is difficult to determine this in terms of the degrees of freedom of the constants $c_{ij}$, we calculate this for the constants
 $a_{j\ell}$ which appear in (\ref{QQ}), or  rather in $f_i(z) =z^{-M_i-1}+\sum_{j=0}^{M _i-1} a_{ji}z^{-j-1}$.   Note that, the corresponding tau-function does not change if we use  Gauss elimination, i.e., if we add a 
multiple of the function $f_i(z)$ to the function $f_j(z)$
With this we can eliminate  
with $f_i(z)$ the constant  $a_{M_i,j}$ in $f_j(z)$ for all $j<i$. This eliminates all dependence  in the constants $a_{j\ell}$ and no more constants can be set to zero. 
Hence, the degrees of freedom that remain are $\lambda_k=M_{k-1}$ for $f_{{k-1}}(z)$,  $\lambda_{k-1}=M_{k-2}-1$ for $f_{{k-1}}(z)$, $\ldots$,
 $\lambda_1=M_0-k+1$ for $f_{0}$. If we add this all up,  we obtain $|\lambda |=-\frac12k(k-1)+\sum_{i=0}^{k-1} M_i$, the desired result.

Now recall that the set of all polynomial tau-functions of the KP hierarchy is the orbit
${\cal O}_0$ of $\mathbb{C} 1\in B$ under the representation $\sigma R\sigma^{-1}$ of the group $GL_\infty$.
Let $P$ be the stabilizer of the line $\mathbb{C} 1$,
let $W$ be the subgroup of permutations of basis vectors of $\mathbb{C}^\infty$ and let $W_0$ be its subgroup, consisting of permutations, permuting vectors with non-positive indices between themselves. Then one has the Bruhat decomposition:
\[ GL_\infty = \bigcup_{w\in W/W_0} UwP \,\,\,(\hbox{disjoint union}).\]
Applying this to $\mathbb{C} 1$, we obtain that the projectivised orbit
$\mathbb{P} {\cal O}_0$ is a disjoint union of Schubert cells $C_w =U w\cdot 1$, $w\in W/W_0$.
It is well known (see, e.g. \cite{KP}) that each $w\cdot 1$ is a Schur polynomial
$s_{\lambda}$ for some partition $\lambda=\lambda (w)$, and the corresponding Schubert cell $C_\lambda =U\cdot s_{\lambda(w)} $ is an affine algebraic variety isomorphic to $\mathbb{C}^
{|\lambda|}$.

On the other hand, by the previous discussion, we have constructed an injective polynomial map from the space $\mathbb{C}^{|\lambda|}$
to the Schubert cell $C_{\lambda}$.
But, by Nagata's lemma, if an affine variety X is embedded in an irreducible
affine variety Y of the same dimension, then either $X=Y$, or the complement $Z$ of $X$ in $Y$ is a closed subvariety of $Y$ of codimension $1$. Since in our situation $Y$ is an affine space, there exists a polynomial $F$ on $Y$, whose set of zeros is $Z$. But then the restriction of $F$ to $X$
is a non-constant invertible polynomial function on $X$, which in our situation is
an affine space as well. This is a contradiction.
\\
\ \\
\noindent {\bf (b)} By part  (a), every $\tau_n$ must be of the form (\ref{Pt}).
Since  we can shift the index $n$ of $\tau_n$, we may assume, without loss of generality,  that $n=k$ and that $\tau_k(t)= \tau_{\lambda_1,\lambda_2,\ldots,\lambda_k}(t;c_1,c_2,\ldots ,c_k)$. Since (\ref{eqP6}) and (\ref{Pt}) give the same tau-function, we find that
\[
\tau_k(t)=W(s_{\lambda_1+k-1}(t+c_{1}), s_{\lambda_2+k-2}(t+c_{2}),\ldots,s_{\lambda_k}(t+c_{k})).
\] 
Using the relation between MKP tau-functions and the infinite flag manifold, as used in  \cite{KP} and \cite{HL},
see also Remark  \ref{R}, we have
\[
\sigma^{-1}(\tau_k)=w_k\wedge w_{k-1}\wedge\cdots\wedge w_1\wedge |0\rangle
\]
and
\[
\sigma^{-1}(\tau_{k+1})=w_{k+1}\wedge w_k\wedge w_{k-1}\wedge\cdots\wedge w_1\wedge |0\rangle ,
\]
hence
the non-zero polynomial tau-function $\tau_{k+1}(t)$ must be the Wronskian determinant of the same functions, but now with one  eigenfunction of $L=\partial$ added. Such an eigenfunction is of the form (\ref{QQ}), thus
\[
\tau_{k+1}(t)=W(s_M(t+d),s_{\lambda_1+k-1}(t+c_{1}), s_{\lambda_2+k-2}(t+c_{2}),\ldots,s_{\lambda_k}(t+c_{k})).
\]
Moreover, we  may assume that $M\not =\lambda_i+k-i$, otherwise we can use Gauss elimination to get a smaller $M$. Now reorder $M,\  \lambda_1+k-1,\ \lambda_2+k-2,\ldots, \ \lambda_k$ to a decreasing order. If $M>\lambda_1+k-1$, then the Wronskian determinant is equal to the first possibility, where $\mu=M-k$. If $\lambda_i+k-i>M>\lambda_{i+1}+k-i-1$ or $\lambda_k>M\not = 0$, we get the second possibility with $\mu=M+i-k$. And finally, when $M=0$, we obtain the last possibility.
\hfill$\square$

\section{Reduction of MKP  to $n$-MKdV}
\label{S5}
Let $n$ be an integer, $n\geq 2$. The $n$-th Gelfand-Dickey hierarchy,  or $n$-KdV, describes the group orbit in a projective representation of the loop group of $SL_n$.
This is not a subgroup of $Gl_\infty$, one has to take a 
bigger group, containing it, as, e.g in \cite{KP}. Then the representation $R$ of
$GL_\infty$ extends to a projective representation, denoted by $\hat{R}$, of this bigger group.
An element of the loop group of $SL_n$ commutes with the operator $q^n$  (in the space $B$), which means that $\tau_{k+n}(t)=\tau_k(t)$ and hence $v_{k+n}(t)=v_k(t)$ and  
$P^\pm_{n+k}(t, \partial)=P^\pm_{k}(t, \partial)$.
This gives that $L_{k+n}=L_{k}$ and that 
\[
\left(L_k^n\right)_-=\left(P_{k}^+\circ \partial^n\circ P_{k}^{+-1}\right)_-=\left(P_{n+k}^+\circ \partial^n\circ P_{k}^{+-1}\right)_-=0\, ,
\]
which means that $L_k^n$ is a differential operator.
Using the Sato-Wilson equations (\ref{Sato}), we deduce that $\frac{\partial P_{k}^+}{\partial t_{jn}}=0$,
for $ j=1,2,\ldots$, and hence, since $L_k=P_{k}^+\circ \partial^n\circ P_{k}^{+-1}$, that also $\frac{\partial L_{k}}{\partial t_{jn}}=0$. The corresponding tau-function then 
satisfies $\frac{\partial \tau_k}{\partial t_{jn}}=a_j \tau_k$ for some constants $a_j$, and hence is  of the form
\begin{equation}
\label{GDtau}
\tau_k(t)=T_k(t) \exp\left( \sum_{j=1}^\infty a_jt_{jn}\right),\quad\text{where } \frac{\partial T_k(t)}{\partial t_{jn}}=0\ \mbox{for } j=1,2,\dots .
\end{equation}
Differentiating (\ref{modKP3}) by $t_{jn}$ and using that $\frac{\partial P_{k}^+}{\partial t_{jn}}=0$, we obtain
\\
\
\\
{\bf The first  formulation of the $n$-MKdV: }
{\it 
\begin{equation}
\label{modnKdV}
{\rm Res}\,   z^{jn+k-\ell}\tau_k(t-[z^{-1}])\tau_\ell(y+[z^{-1}]) \exp\left(\sum_{i=1}^\infty (t_i-y_i)z^i\right)dz=0,
\end{equation}
for all $0\le k,\ell\le n-1$ and  $ j\ge 0$ , provided  that $ jn+k-\ell\ge 0$.
}	
\\
\
\\
Let $\epsilon=\exp\frac{2\pi i}n$. One can reformulate (\ref{modnKdV}) to one identity for each pair $k$ and $\ell$ as in \cite{Giv}, equation (8):
\[
z^{-1}\sum_{a=1}^n (\epsilon^az)^{k-\ell+1+\delta n}\tau_k(t-[(\epsilon^az)^{-1}])\tau_\ell(y+[(\epsilon^az)^{-1}]) \exp\left(\sum_{i=1}^\infty (t_i-y_i)(\epsilon^az)^i\right)
\]
has no negative powers of  $z$,
for $0\le k,\ell\le n-1$, and  $\delta=0$ if $k-\ell \ge 0$ and $=1$ if $k-\ell<0$.

The fact  $P^+_{n}=P_0^+$ and  that $L_0^n$ is a differential operator, gives that    $L_0$ is the $n$-th root of a differential operator \cite{D}, \cite{Dbook}
\[
\begin{aligned}
{\cal L}_0&=\partial^n+ w_{n-2}(t)\partial^{n-2}+\cdots  + w_{1}(t)\partial+ w_{0}(t)=L_0^n=P_n^+(t)\circ \partial^n\circ P_0^+(t)^{-1}\\
&=(\partial +v_{n-1}(t))\circ (\partial+v_{n-2}(t))\circ\cdots\circ(\partial+v_{0}(t))  P_0^+(t)P_0^+(t)^{-1}    \\
&=(\partial +v_{n-1}(t))\circ (\partial+v_{n-2}(t))\circ \cdots\circ (\partial+v_{0}(t))
\, .
\end{aligned}
\]
The explicit form  (\ref{Pll}) of the $v_j(t)$  expressed in terms of the tau-functions (\ref{GDtau}), gives that 
\[
v_0(t)+v_1(t)+\cdots v_{n-1}(t)=0,\quad  \text{and that } \frac{\partial v_k(t)}{\partial t_{jn}}=0,\ \text{for all }j=1,2,\dots  .
\]
Note that by (\ref{LLL}):
\[
{\cal L}_j:= L_j^n=(\partial +v_{j-1}(t))\circ (\partial+v_{j-2}(t))\circ \cdots\circ (\partial+v_{0}(t))\circ (\partial +v_{n-1}(t))\circ (\partial+v_{n-2}(t))\circ \cdots\circ(\partial+v_{j}(t))\, ,
\]
which is a Darboux transformation of ${\cal L}_0$, i.e. a cyclic permutation of the factors $\partial+v_j$ of ${\cal L}_0$.

Since now $L_i={\cal L}_i^{\frac{1}{n}}$ is only expressed in the $v_j$, the second set of equations of (\ref{Dickey}), which now have the form
\begin{equation}
\label{Diki}
\frac{\partial v_i}{\partial t_j}=\left({\cal L}_{i+1}^{\frac{j}{n}}\right)_+\circ (\partial +v_i)-(\partial +v_i)\circ \left({\cal L}_{i}^{\frac{j}{n}}\right)_+,\quad\text{where } {\cal L}_{n+i}={\cal L}_{i}, 
\end{equation}
imply the first ones, the Lax equations, of (\ref{Dickey}).

We can reformulate the equations (\ref{Diki})  by one  compact formula (see e.g. \cite{F}).
Let 
\begin{equation}
\label{L}
{\cal L}={\rm diag}\,\left( {\cal L}_0,{\cal L}_1,\dots,{\cal L}_{n-1}\right) 
\end{equation}
and 
\begin{equation}
\label{M}
M=\begin{pmatrix}0&\cdots&\cdots&0 &\partial+v_{n-1}(t)
\\
\partial+v_0(t)&0&&&0
\\
0&\partial+v_{1}(t)&0&&\vdots
\\
\vdots&\ddots&\ddots &\ddots&\vdots
\\
0&\cdots&0&\partial +v_{n-2}(t)&0
\end{pmatrix}
\end{equation}
Then 
$
{\cal L}=M^n,
$
and the equation (\ref{Diki}) is exactly the $(i+2)\mod n$-th row of the equation
\begin{equation}
\label{refDickey}
	\frac{\partial M}{\partial t_j}=\left[ \left({\cal L}^{\frac{j}{n}}\right)_+,M\right]\,,\,\, j=1,2,\cdots .
\end{equation}
Hence we obtain:\\
\
\\
{\bf The second  formulation of the $n$-MKdV: }
\\
  {\it Let $U_n=\mathbb{C}[ v_i^{(m)}| i=0, 1,2,\cdots,  n-1,\  m\in\mathbb{Z}_{\ge 0}]/(v_0+v_1+ \cdots v_{n-1})$ be the quotient of the algebra of differential polynomials in  
 $v_j$  by the differential ideal, generated by $v_0+v_1 +\cdots +v_{n-1}$. Then  the $n$-MKdV hierarchy is the system of evolution equations
    (\ref{refDickey}) in $U_n$, where $\cal L$ and $M $ are given by (\ref{L}) and (\ref{M}).
}
\\
\ 
\\
{\it Example.} For $n=2$, we get the modified KdV equation in $v=v_0=-v_1$. Indeed:
\[
\begin{aligned}
{\cal L}_0&=\partial^2 +u_0=(\partial-v)\circ (\partial +v)=\partial^2 +\frac{\partial v}{\partial t_1}-v^2\, ,\\
{\cal L}_1&=\partial^2 +u_1=(\partial+v)\circ (\partial -v)=\partial^2 -\frac{\partial v}{\partial t_1}-v^2\, ,\\
\end{aligned}
\]
and 
\[
\frac{\partial v}{\partial t_j}=\left({\cal L}_1^{\frac{j}2}\right)_+\circ (\partial +v)-(\partial +v)\circ \left({\cal L}_0^{\frac{j}2}\right)_+, \quad j=1,3,5,\ldots .
\]
For $j=3$ this gives the classical modified KdV equation:
\[
\frac{\partial v}{\partial t_3}=-\frac32 v^2\frac{\partial v}{\partial t_1}+\frac{\partial^3 v}{\partial t_1^3}\, .
\]

\section{Polynomial solutions of $n$-KdV  and $n$-MKdV}
\label{S8}
We can use the ideas of  Section \ref{S7} to obtain polynomial tau-functions of $n$-MKdV. We will first construct a polynomial tau-function for the $n$-KdV hierarchy. 
Let $\pi$ be a permutation of $1,2,\ldots, n$, such that $\pi(i)=j_i$, and choose $n$ formal power series 
\[
f_i(z)=z^{j_i-1}+\sum_{k=j_i}^\infty a_{ki} z^k,\quad i=1,2,\ldots, n.
\]
Choose non-negative integers $m_1,m_2,\ldots ,m_n$, such that at least one $m_i=0$ and one $m_i$ non-zero (all $m_i =0$  would lead to 
the trivial solution $\tau_0=1$). We construct $L_0=\partial$ eigenfunctions from these data.
For $\ell=1,2,\ldots m_i$, define
\begin{equation}\label{qqq}
q_{\ell,i}(t)= {\rm Res} z^{-\ell n}f_i(z) e^{t\cdot z}dz =s_{\ell n-j_i}(t)+\sum_{k\ge j_i}a_{ki} s_{\ell n-k-1}(t)= s_{\ell n-j_i}(t+c_{i})\, ,
\end{equation}
for certain constants $c_{i}= (c_{1i},c_{2i}, \ldots)$.
Then $\tau_0(t)$ is the Wronskian determinant  of
all functions 
\[
s_{\ell n-j_i}(t+c_{i}),\quad\mbox{for }1\le i\le n,\ 1\le \ell \le m_i\ \mbox{and } \ell n-j_i\ge 0 \, .
\]
This determinant clearly becomes zero after differentiating by $t_{pn}$ since differentiating the function $s_{\ell n-j_i}(t+c_{i})$ by $t_{pn}$ gives 
$s_{(\ell-p) n-j_i}(t+c_{i})$, which is either zero if $(\ell-p) n-j_i<0$ or
it already appears as an eigenfunction in the Wronskian determinant. Hence $\tau_0(t)$ is an $n$-KdV tau-function.

We obtain $\tau_1$ by adding  the eigenfunction $s_{(m_1+1)n-j_1}(t+c_{1})$ to the Wronskian determinant. We obtain $\tau_2$ by adding 
this function and also $s_{(m_2+1)n-j_2}(t+c_{2})$.
For $\tau_3$ we add besides these two also $s_{(m_3+1)n-j_3}(t+c_{3})$, etc.
For $\tau _n$ we add the functions
\[
s_{(m_1+1)n-j_1}(t+c_{1}), s_{(m_2+1)n-j_2}(t+c_{2}),\ldots,s_{(m_n+1)n-j_n}(t+c_{n})\, .
\]
This however gives no new tau-function:  it is straightforward to check, but rather tedious, that $\tau_n$ is a scalar multiple of $\tau_0$.
In fact the theorem, that we shall prove later on in this section, then implies that this construction gives  all possible polynomial tau-functions for $n$-MKdV.

\begin{example}
Let us inspect the case $n=2$. In this case either  $m_1=0$ or $m_2=0$ and $\pi$ is the identity or the transposition $(12)$.  This gives two possible solutions, viz
\[
\begin{aligned}
  &\tau_0(t)=s_{k,k-1,\ldots, 2,1}(t+c)&\mbox{and }&\tau_1(t)=
  s_{k+1,k,k-1,\ldots, 2,1}(t+c), \ \mbox{or}\\
  &\tau_0(t)=s_{k,k-1,\ldots, 2,1}(t+c)&\mbox{and }&\tau_1(t)=s_{k-1,k-2,\ldots, 2,1}
  (t+c)\, ,
\end{aligned}
\]
where $c=(c_1, c_2,...)$, which are all polynomial tau-functions of the KdV and the modified KdV hierarches. This is a result of \cite{KP}, Theorem 9.1(b). Note that these tau-functions are independent of the even times $t_{2k}$.
\end{example}

For general $n$ to describe all tau-functions that satisfy the $n$-MKdV hierarchy in terms of a formula like (\ref{Pt}) is rather complicated. Not only are  there special partitions $\lambda$ connected to the case of $n$-KdV. But also
instead of arbitrary constants $c_i=(c_{1i},c_{2i},\ldots)$ connected to part $\lambda_i$ of the partition $\lambda$, there are certain restrictions. This time there are  series of constants that depend on the shifted parts $\lambda_i-i+1$, but then calculated modulo $n$. Hence, there are $n$ of such series $c_{\overline i}=(c_{1\overline i},c_{2\overline i},\dots)$ of which at most $n-1$ appear in the tau-function. Here and thereafter $\bar{s}$  stands for remander of the division of $s$ by $n$.

We claim that 
the Wronskian determinant 
\begin{equation}
\label{qqqq}
W(s_{\lambda_1+k-1}(t+c_{\overline{\lambda_1}}),s_{\lambda_2+k-2}(t+c_{\overline{\lambda_2-1}}),\ldots,s_{\lambda_k}(t+c_{\overline{\lambda_k-k+1}} )
)\, ,
\end{equation}
is a polynomial tau-function of the $n$-KdV if and only if the set of shifted parts
\[
V_\lambda=\{ \lambda_1,\lambda_2-1,\lambda_3-2,\ldots,\lambda_k-k+1,-k,-k-1,-k-2,\ldots\}
\]
satisfies the condition that 
\[
\mbox{if }j\in V_\lambda,\ \mbox{ then also }j-n\in V_\lambda\, .
\]
This condition reflects the condition that if the eigenfunction $q_{\ell,i}(t)$,  defined in (\ref{qqq}) appears in the Wronskian determinant, then also $q_{\ell-n,i}(t)$, if it is non-zero, must appear in this determinant as well. Or stated differently, if $s_{\lambda_i+k-i}(t+c_{\overline{\lambda_i-i+1}})$  appears in the Wronskian determinant of (\ref{qqqq}), then either
$\frac{\partial s_{\lambda_i+k-i}(t+c_{\overline{\lambda_i-i+1} } )}{\partial t_n}=0$ or $ s_{\lambda_i+k-i-n}(t+c_{\overline{\lambda_i-i+1}})$ also appears in this determinant as well.
This leads us to the following notion.
\begin{definition}
  A partition $\lambda$ is called $n$-periodic if the corresponding infinite sequence $V_\lambda$
  is mapped to itself when subtracting $n$ from each term.
\end{definition}

\begin{theorem}
\label{TKdV}
All polynomial tau-functions of the $n$-KdV hierarchy are, up to a constant factor, of the form
\begin{equation}
\label{PtKdV}
\tau^n_{\lambda_1,\lambda_2,\ldots,\lambda_k}(t;c_{\overline{\lambda_1}},c_{\overline{\lambda_2-1}},\ldots ,c_{\overline{\lambda_k-k+1}})=\det \left( s_{\lambda_i+j-i}(t_1+c_{1,\overline{\lambda_i-i+1}},t_2+c_{2,\overline{\lambda_i-i+1}}\ldots)\right)_{1\le i,j\le k}\, ,
\end{equation}
where $\lambda=(\lambda_1,\lambda_2, \ldots,\lambda_k)$ is an $n$-periodic partition. Here  the $c_{\overline i}=(c_{1\overline i},c_{2\overline i},\dots)$ for $i=1,2,\ldots n$ (where at most $n-1$ of such  $\overline i$'s appear)
are arbitrary constants.
\end{theorem}
\
\\
Before we give the proof, let us make calculations in an explicit example.
Let $n=4$ and $\lambda=(6,3,2,1)$. Then
\[
V_\lambda =\{ 6,2,0,-2,-4,-5,-6,\ldots\}\,, 
\]
hence $\lambda$ is 4-periodic, and the corresponding tau-function is 
\begin{equation}
\label{tautau}
\begin{aligned}
\tau_{(6,3,2,1)}^4(t; c_{\overline2},c_{\overline2}, c_{\overline 4},c_{\overline 2})=&
W(s_9(t+c_{\overline 2}),s_5(t+c_{\overline 2}),s_3(t+c_{\overline 2}),s_1(t+c_{\overline 2}))
\\
=&
\left|
\begin{matrix}s_6(t+c_{\overline 2})&s_7(t+c_{\overline 2})&s_8(t+c_{\overline 2})&s_9(t+c_{\overline 2})\\
s_2(t+c_{\overline 2})&s_3(t+c_{\overline 2})&s_4(t+c_{\overline 2})&s_5(t+c_{\overline 2})\\
s_0(t+c_{\overline 4})&s_1(t+c_{\overline 4})&s_2(t+c_{\overline 4})&s_3(t+c_{\overline 4})\\
0&0&s_0(t+c_{\overline 2})&s_1(t+c_{\overline 2})\\
\end{matrix}
\right|\, ,
\end{aligned}
\end{equation}
which depends on two series of constants, viz. $c_{\overline 6}=c_{\overline2}$ and $c_{\overline 0}=c_{\overline 4}$.
The $\overline 6$ and $\overline 0$ are the elements of the following set 
\[
U^{(4)}_\lambda=\{6,2,0,-2\}\backslash\{2,-2,-4,-6\}=\{6,0\},
\]
which are all the elements $j$ of $V_\lambda$ where one removes all elements   $ j-4$. 

Now 
\[
\begin{aligned}
&s_9(t+c_{\overline 2})= s_{9}(t)+\sum_{j=0}^{8} 
a_{9-j,\overline{2}}s_j(t)\ \mbox{and }
& f_6(z)=z^{-10}+\sum_{j=0}^{8} 
a_{9-j,\overline{2}}z^{-j-1},\\
&s_5(t+c_{\overline 2})=s_5(t)+\sum_{j=0}^{4}a_{5-j,\overline{2}}s_j(t),\ 
& f_2(z)=z^{-6}+\sum_{j=0}^{4} 
a_{5-j,\overline{2}}z^{-j-1},\\
&s_3(t+c_{\overline 4})=s_3(t)+\sum_{j=0}^{2}a_{3-j,\overline{4}}s_j(t),\ 
 &f_0(z)=z^{-4}+\sum_{j=0}^{2} 
a_{3-j,\overline{4}}z^{-j-1},\\
&s_1(t+c_{\overline 2})=s_1(t)+ a_{1,\overline{2}}s_0(t), \ 
&f_{-2}(z)=z^{-2}+ a_{1,\overline{2}}z^{-1},\\
\end{aligned}
\]
where $a_{k,\overline j}=s_k(c_{\overline j})$.
And as in the proof of Theorem \ref{T7} we can eliminate the coefficients of $z^{-2}$,  in $f_0(z)$, $f_2(z)$ and  $f_6(z)$, and the coefficient of  $z^{-4}$  in  $f_2(z)$ and  $f_6(z)$ and the coefficient of $z^{-6}$ in $f_6(z)$, leaving  a freedom of $9-3=6=\lambda_1$  in $f_6(z)$ and similarly  a freedom of $3-1=2=\lambda_3$ in $f_0(z)$. Hence the dimension of the space of polynomials  (\ref{tautau}) is 
\[
8=6+2=\lambda_1+\lambda_3=\sum_{\lambda_i\in \Lambda^{(4)}(\lambda)} \lambda_i\, ,
\]
where 
\[ \Lambda^{(4)}((6,3,2,1)=\{\lambda_1,\lambda_3\}=\{6,2\}\, .
\]
Let us next investigate the element $s_{(6,3,2,1)}(t)$ the corresponding element under $\sigma^{-1}$ is
\[
\begin{aligned}
\sigma^{-1}(s_{(6,3,2,1)}(t))=& e_6\wedge e_2\wedge e_0\wedge e_{-2} \wedge e_{-4}\wedge e_{-5}\wedge \cdots\\
=&t^{-1}u_2\wedge u_2 \wedge tu_4\wedge tu_2\wedge t^{2}u_4\wedge t^{2}u_3\wedge t^{2}u_2\wedge t^{2}u_1\wedge t^{3}u_4\wedge\cdots\, .
\end{aligned}
\]
Here we make  the identification $t^{-k}u_j=e_{4k+j}$ and $t^ke_{ij}=\sum_{s\in\mathbb{Z}}E_{4(s-k)+i,4s+j}$ as in \cite{KP} , eq. (9.1-2).
And this is up to some infinite reordering "equal to"
\[
\begin{aligned}
&\hat R(t^{1}e_{11}+t^{-2}e_{22}+t^{1}e_{33}+e_{44})(t^{}u_4\wedge t^{}u_3\wedge t^{}u_2\wedge t^{}u_1\wedge t^{2}u_4\wedge\cdots ) \\
&\quad= \hat R(t^{1}e_{11}+t^{-2}e_{22}+t^{1}e_{33}+e_{44})|0\rangle\, .
\end{aligned}
\]
We now reconstruct our $\lambda$ from the element $t^{1}e_{11}+t^{-2}e_{22}+t^{1}e_{33}+e_{44}$. For this we invert the process above. We first calculate the corresponding infinite wedge product and need to find the place of $e_6=t^{-1}u_2=t^{-2}e_{22}tu_2$ and $e_0=tu_4=e_{44}tu_4$ in this product. It  is the place $0$ and the
place $-2$, which gives the elements $\lambda_1=6-0$ and $\lambda_3=0-({-2})=2$ of $\lambda$ .

We now want to use some of the above features of the example in the\\
\ \\ 
{\bf Proof of Theorem \ref{TKdV}. }
First observe that (\ref{qqqq}) is equal to (\ref{PtKdV}).
 
As in the proof of Theorem \ref{T7}, we can calculate the degrees of freedom of the constants in a similar way. Let $\lambda=(\lambda_1,...,\lambda_k)$ be a partition. As before,
\[
s_{\lambda_i+k-i}(t+c_{\overline{\lambda_i-i+1}})=s_{\lambda_i+k-i}(t)+\sum_{j=0}^{\lambda_i+k-i-1} 
a_{\lambda_i+k-i-j,\overline{\lambda_i-i+1}}
 s_j(t)= \mbox{Res}\, f_{\lambda_i-i+1}(z) e^{t\cdot z} dz\, ,
\]
for $a_{j,\overline{\lambda_i-i+1}}=s_j(c_{\overline{\lambda_i-i+1}})$, and
\[ 
f_{\lambda_i-i+1}(z)=z^{-(\lambda_i-i+1)-k}+\sum_{j=0}^{\lambda_i+k-i-1} a_{j,\overline{\lambda_i-i+1}}z^{-(\lambda_i+k-i)+j-1}\, .
\]
Note that $f_{\lambda_i-i+1-n}(z)$  also appears as some  $z^nf_{\lambda_j-j+1}(z)$, for some $j>i$ and it has the form
\[
f_{\lambda_i-i+1-n}(z)= (z^n f_{\lambda_i-i+1}(z))_-=z^{-(\lambda_i-i+1)-k-n}+\sum_{j=0}^{\lambda_i+k-i-1-n} a_{j,\overline{\lambda_i-i+1}}z^{-(\lambda_i+k-i)+j+n-1}\, .
\]
Hence, proceeding in a similar way as in the proof of Theorem \ref{T7}, we can use $f_{\lambda_i-i+1}(z)$ to eliminate the constant $a_{\lambda_\ell-\lambda_i+i-\ell-1,\overline{\lambda_\ell-\ell+1}}$,  in front of $z^{-(\lambda_i-i+1)-k}$ in $ f_{\lambda_\ell-\ell+1}(z)$ for all $\ell<i$. Note that we cannot eliminate more constants.  Hence we have  $\lambda_k$ degrees of freedom for $f_{\lambda_k-k+1}(z)$, $\lambda_{k-1}$ for $f_{\lambda_{k-1}-k+2}(z)$,  $\lambda_{k-2}$ for $f_{\lambda_{k-2}-k+3}(z)$,  $\dots$, $\lambda_1$ for $f_{\lambda_1}(z)$. This is similar to the KP case,
except that some of the $f_{\lambda_i-i+1}(z)$ are related, as described above. Hence we have to find those $f_{\lambda_i-i+1}(z)$ with the highest possible index
that are not related to the one with a higher index. These are all the $f_j(z)$'s, with $j$ from the following set:
\[
U^{(n)}_\lambda=\{\lambda_1, \lambda_2-1, \ldots,\lambda_k-k+1\}
\backslash \{\lambda_1-n, \lambda_2-n+1, \ldots,\lambda_k-n-k+1\}\,.
\]
If $j\in U_\lambda$, then $j=\lambda_i-i+1$ for some $i$ and $f_j(z)=f_{\lambda_i-i+1}(z)$ has $\lambda_i$ degrees of freedom. Hence, defining 
\[
\Lambda^{(n)}(\lambda)=\{\lambda_i|\lambda_i-i+1\in U^{(n)}_\lambda\}\, ,
\]
the freedom of choosing constants (or the dimension of this subspace of polynomials) is equal to 
\[
\sum_{\lambda_i\in \Lambda^{(n)}(\lambda)} \lambda_i\, . 
\]

As before, the the tau-function (\ref {PtKdV})
is the image under  $\sigma$ in $B$  of the following element of $F^{(0)}$:
\begin{equation}
\label{xxx}
\begin{aligned}
\left(e_{\lambda_1}+\sum_{j=1}^{\lambda_1+k-1} a_{j-1, \overline{\lambda_1}}e_{\lambda_1-j}
\right)\wedge\left(e_{\lambda_2-1}+\sum_{j=1}^{\lambda_2+k-2} a_{j-1, \overline{\lambda_2-1}}e_{\lambda_2-1-j}
\right)\wedge\cdots\\
\cdots\wedge 
\left(e_{\lambda_k-k+1}+\sum_{j=1}^{\lambda_k} a_{j-1, \overline{\lambda_k-k+1}}e_{\lambda_k-k+1-j}
\right)\wedge e_{-k}\wedge e_{-k-1}\wedge\cdots\, ,
\end{aligned}
\end{equation}
which is equal to
\[
R\left(I+\sum_{i=1}^k\sum_{j=1}^{\lambda_i+k-i} a_{j-1, \overline{\lambda_i-i+1}}E_{\lambda_i-i+1-j,\lambda_i-i+1 }\right)
(e_{\lambda_1}\wedge e_{\lambda_2-1}\wedge \cdots\wedge e_{\lambda_k-k+1}\wedge e_{-k}\wedge e_{-k-1}\wedge\cdots)\, ,
\]
where 
\[
\sigma(e_{\lambda_1}\wedge e_{\lambda_2-1}\wedge \cdots\wedge e_{\lambda_k-k+1}\wedge e_{-k}\wedge e_{-k-1}\wedge\cdots)=s_\lambda(t)\, .
\]
We can rewrite (\ref{xxx}) as follows:
\[
R\left(I+\sum_{p\in U_\lambda^{(n)}}  \sum_{0\le s<\frac{k+p}{n}}^{} \sum_{j=1}^{p+k-sn-1} a_{j-1, \overline{p}}E_{p-j-sn,p-sn }\right)
(e_{\lambda_1}\wedge e_{\lambda_2-1}\wedge \cdots\wedge e_{\lambda_k-k+1}\wedge e_{-k}\wedge e_{-k-1}\wedge\cdots)\, .
\]
Note that replacing the upper bound $p+k-sn-1$ of $j$ by $p+k-1$ does not change the element. We can also drop the lower bound of $s$ because this will give a matrix element that acts as zero on every vector of the wedge product $\sigma^{-1}(s_\lambda(t))$. We can also drop the upper bound of $s$. Indeed, if we do that, the new element transforms the element $e_\ell$ for $\ell\le -k$ into
an element of the form $v_\ell=e_\ell+\sum_{-\infty<<i<\ell} b_i e_i$. We can then use the $v_j$ for $j<\ell$ to eliminate all the coefficients of $b_i$ (we have to do this procedure infinitely many times).
In this way we get that (\ref{xxx}) is equal  to
\[
\hat R\left(I+\sum_{p\in U_\lambda^{(n)}}  \sum_{j=1}^{p+k-1} a_{j-1, \overline{p}} \sum_{s\in\mathbb{Z}}E_{ p-j+sn, p+sn }\right)
(e_{\lambda_1}\wedge e_{\lambda_2-1}\wedge \cdots\wedge e_{\lambda_k-k+1}\wedge e_{-k}\wedge e_{-k-1}\wedge\cdots)\, .
\]

Now we relate the above element of the completed $GL_\infty$ to an element of
of the loop group $SL_n(\mathbb{C}[t, t^{-1}])$ by making the identification $t^{-k}u_j=e_{kn+j}$ and $t^ke_{ij}=\sum_{s\in\mathbb{Z}}E_{(s-k)n+i,sn+j}$ as in \cite{KP}, eq. (9.1-2).
Let $$U=\{ A(t)\in SL_n(\mathbb{C}[t])|\, A(0)\ \mbox{is upper triangular with 1's }\mbox{on the diagonal}\}\, .$$
Then, under the above identification we have
\[
I+\sum_{p\in U_\lambda^{(n)}}  \sum_{j=1}^{p+k-1} a_{j-1, \overline{p}} \sum_{s\in\mathbb{Z}}E_{ p-j+sn, p+sn }\in U.
\]

Let $ T=\{\sum_{i=1}^n t^{k_i}e_{ii}
|\, k_i\in \mathbb{Z},\,\sum_{i=1}^n k_i=0\} \subset SL_n(\mathbb{C}[t,t^{-1})\}$.
  Fix $ w=\sum_{i=1}^n t^{k_i}e_{ii}\in T$. We want to find the partition that corresponds to  
$\hat R(w)|0\rangle$, i.e., to find $\lambda$ such that $\sigma(\hat R(w)|0\rangle)=s_\lambda(t)$. In fact,  if $\lambda=(\lambda_1,\lambda_2,\ldots )$,  we want to find its parts $\lambda_i$ that are in $\Lambda^{(n)}(\lambda)$. We will denote these elements by $\hat\lambda_1, \hat\lambda_2,\ldots,\hat\lambda_p$. Now, $\hat R(w)|0\rangle$ is a semi-infinte wedge product of the elements $t^{k_i+j}u_i=e_{-(k_i+j)n+i}$, for $j>0$ and all $1\le i\le n$. We have to order these $e_\ell$ in a decreasing order in this wedge product, from which we then can determine the corresponding partition $\lambda$.  For this, first reorder the elements $k_i$ to the decreasing order  without interchanging $k_i$'s, if they are the same. Then  $p$ is the same as the number of $k_i$'s which are smaller than the maximum of this set.
Let $\pi$ be the permutation that assigns  to $i$ the number $j$ if $k_j$ is in the $i$-th place in the decreasing order.
The corresponding $\Lambda^{(n)}(\lambda)$ has $p$ elements $\hat\lambda_i$, which we put in decreasing order: $\hat\lambda_1\ge \hat\lambda_2\ge \cdots\ge \hat\lambda_p$.
The part $\lambda_1$, which is always an element of $\Lambda^n(\lambda)$,  corresponds to the place of $t^{k_{\pi(n)}+1}u_{\pi(n)}=e_{-k_{\pi(n)}n+\pi(n)-n}$ in the semi-infinite wedge
product, which is always on the $0$-th place. Hence 
\[
\hat\lambda_1=\lambda_1=-k_{\pi(n)}n+\pi(n)-n\, 
\] 
and $\lambda_2=-k_{\pi(n)}n+\pi(n)-2n+1$, since it corresponds to  $t^{k_{\pi(n)}+2}u_{\pi(n)}=e_{-k_{\pi(n)}n+\pi(n)-2n}$, then $ \lambda_3=-k_{\pi(n)}n+\pi(n)-3n+2$ and we continue  as long as $k_{\pi(n)}+1$, $k_{\pi(n)}+2,\dots$ is smaller than $k_{\pi(n-1)}$.  
To determine  $\hat\lambda_2$ of $\Lambda^{(n)}(\lambda)$, is already a bit more complicated. One has to consider two cases.
It is $\lambda_{k_{\pi(n-1)}-k_{\pi(n)}+2}$, if $k_{\pi(n-1)}=k_{\pi(n)}$ or if $k_{\pi(n-1)}>k_{\pi(n)}$ and $\pi(n-1)<\pi(n)$. Then the element
$t^{k_{\pi(n-1)}+1}u_{\pi(n-1)}$, which is equal to
 $e_{-k_{\pi(n-1)}n-+{\pi(n-1)}-n}$ is in the $-k_{\pi(n-1)}+k_{\pi(n)}-1$-th place in the semi-infinite wedge product. Hence $\hat\lambda_2=-k_{\pi(n-1)}(n-1)-k_{\pi(n)}+{\pi(n-1)}-(n-1)$. 
However, if $k_{\pi(n-1)}>k_{\pi(n)}$ and $\pi(n-1)>\pi(n)$, then $\hat\lambda_2=\lambda_{k_{\pi(n-1)}-k_{\pi(n)}+1}$ and this corresponds to the same element $e_{-k_{\pi(n-1)}n+{\pi(n-1)}-n}$, hence $\hat\lambda_2=-k_{\pi(n-1)}(n-1)-k_{\pi(n)}+{\pi(n-1)}-(n-1)-1$. The extra $-1$ at the end comes from the inversion of $\pi$ between the elements $n-1$ and $n$,
viz. in this case $\pi(n-1)>\pi(n)$. The number of inversions will turn out to be important, so let us introduce some notation. Let 
\[
J_j=| \{i>j | \pi(i)<\pi(j)\}|,
\] 
then 
\[
\hat\lambda_2=-k_{\pi(n-1)}(n-1)-k_{\pi(n)}+\pi(n-1)-(n-1)-J_{n-1}\, .
\]
For the next one we have
$\hat\lambda_3=\lambda _{2k_{\pi(n-2)}-k_{\pi(n-1)}-k_{\pi(n)}}+2-J_{n-2}$  and the corresponding element is
$
t^{k_{\pi(n-2)}+1}u_{\pi(n-2)}=
e_{-k_{\pi(n-2)}n+{\pi(n-2)}-n}$,
which gives
\[
\hat\lambda_3=-k_{\pi(n-2)}(n-2)-k_{\pi(n-1)}-k_{\pi(n)}+\pi(n-2)-(n-2)-J_{n-2}\, .
\]
Continuing in this way we find 
\[
\hat\lambda_j=-k_{\pi(n-j+1)}(n-j+1)-k_{\pi(n-j+2)}- \cdots k_{\pi(n-1)}-k_{\pi(n)}+\pi(n-j+1)-(n-j+i)-J_{n-j+1}\, ,
\]
where the last one is $\hat\lambda_p$.
The dimension of this space is  $\hat\lambda_1+\hat\lambda_2+\cdots+\hat\lambda_p$, which is equal to
\begin{equation}
\label{dim1}
\sum_{j=n-p+1}^n (n-p-2j +1)k_{\pi (j)}+\pi(j)-j-J_{j}\, .
\end{equation}
Since $\sum_i k_i=0$,  we can add a multiple of this sum, thus equation (\ref{dim1}) is equal to 
\begin{equation}
\label{dim2}
p\sum_{i=1}^{n-p} k_{\pi(i)}+\sum_{j=n-p+1}^n (n-2j+1)k_{\pi (j)}+\pi(j)-j-J_{j} \, .
\end{equation}
Now, $k_{\pi(1)}=k_{\pi(2)}=\cdots=k_{\pi(n-p)}$, hence 
\[
p\sum_{i=1}^{n-p} k_{\pi(i)}=p(n-p)k_{\pi(1)}=\sum_{i=1}^{n-p} (n-2i+1)k_{\pi(i)}.
\]
Thus (\ref{dim2}) is equal to 
\begin{equation}
\label{dim3}
\sum_{i=1}^n (n-2i+1)k_{\pi (i)}-
\sum_{j=n-p+1}^n j-\pi(j)+J_{j} \, .
\end{equation}
Note that  $\pi(1)<\pi(2)< \cdots < \pi(n-p)$ and $j-\pi(j)+J_j$ are the number of inversions between $j$ and all elements $i$ with $i<j$, thus 
\[
\sum_{j=n-p+1}^n j-\pi(j)+J_{j}=\ \mbox{number of inversions of }\pi\, ,
\]
hence, the dimension of the space which corresponds to $w$ is
\begin{equation}
\label{dimT}
\sum_{\lambda_i\in \Lambda^{(n)}(\lambda)} \lambda_i=\sum_{i=1}^n (n-2i+1)k_{\pi (i)}-(\mbox{number of inversions of }\pi)
\end{equation}

We now have to prove that this is indeed the right dimension to obtain all possible polynomial tau-functions.
Recall that the set of all polynomial tau-functions of the $n$-KdV hierarchy is the orbit
${\cal O}_0^n$ of $\mathbb{C} 1\in B$ under the projecive representation $\hat R$ of the group $SL_n(\mathbb{C}[t,t^{-1}])$.
Let $P=SL_n(\mathbb{C}[t])$.  
Then one has the Bruhat decomposition:
\[ SL_n(\mathbb{C}[t,t^{-1}]) = \bigcup_{w\in T} Uw P\,\, \,(\hbox{disjoint union}).\]
Applying this to $\mathbb{C} 1$, we obtain that the projectivisation of the orbit
${\cal O}_0^n$ is a disjoint union of Schubert cells $C_w =U w\cdot 1$, for all possible $w=\mbox{diag}(t^{k_1},...,t^{k_n})\in T$.
Now, $UwP=w w^{-1}U wP$, hence elements of $U$ that $w$ conjugates to elements in $P$ get absorbed in $P$ , and the elements $t^c e_{ij}\in U$ that get mapped under conjugation by $w$ to elements  $t^de_{ij}$  with $d<0$ give the cell. Hence we have to count the  possible values of $c$ such that $c-k_i+k_j<0$. This is straigtforward, for $i<j$ it is $|k_i-k_j|$ if $k_i>k_j$  and $0$ otherwise. For $j<i$ we find $|k_i-k_j|-1 $ if $k_i>k_j$  and $0$ otherwise. Hence, we obtain as dimension the sum of all values $| k_i-k_j |$ for $1\le i<j\le n$, where we have to subtract 1 if
$k_i>k_j$. We find that the dimension of this Schubert cell is
\[
\sum_{1\le i<j\le n} \left(| k_i-k_j |- \begin{cases} 1&\mbox{ if }k_i>k_j,\\
0,& \mbox{otherwise.}
\end{cases}\right)
\]
Now ordering the $ k_i$'s  in decreasing order (where $\pi$ is the permutation as before), we can remove the absolute value and obtain that the dimension is equal to 
\[
\sum_{1\le i<j\le n} \left(k_{\pi(i)}-k_{\pi(j)} -
\begin{cases} 1&\mbox{ if }\pi(i)>\pi(j)\\
0,& \mbox{otherwise.}
\end{cases}\right) \]
In this sum  $k_{\pi(i)}$ appears $n-1$ times, with $n-i$ plus signs and $i-1$ minus signs, hence we obtain that the dimension of the Schubert cell $C_w$ is equal to
\[
\sum_{i=1}^n (n-2i+1)k_{\pi (i)}-(\mbox{number of inversions of }\pi)=\sum_{\lambda_i\in \Lambda^{(n)}(\lambda)} \lambda_i\, ,
\]
which is the dimension of the space of polynomials of the form (\ref{PtKdV}).
The same algebro-geometric argument as in the KP case completes the proof of  the theorem.\hfill$\square$

\begin{example}
For $n=3$ we have the following possible polynomial  tau-functions of the $3$-KdV hierarchy.
Let $k,\ell=0
,1,2,\ldots$, then we find two series (see (\ref{PtKdV})):
\[
\tau^3_{k+2\ell,k+2\ell -2,\ldots \ell+2,\ell, \underline \ell ,\ell-1,\underline{\ell-1},\cdots, 1,\underline 1 }(t;c,c,\ldots c, c, \underline c, c,\underline c,\ldots ,c,\underline c)
\]
and 
\[
\tau^3_{k+2\ell+1,k+2\ell -1,\ldots\ell+3, \ell+1, \underline \ell ,\ell ,\underline{\ell-1},\ell-1,\cdots, \underline 1 , 1}(t;c,c,\ldots ,c, c, \underline c, c,\underline c,c,\ldots ,\underline c,c)
\]
We have at most two series of constants that appear, viz. $c=(c_1,c_2,c_3,\ldots)$ and $\underline c=(\underline c_1,\underline c_2,\underline c_3,\ldots)$, and $c$ is coupled to the parts of the partition which are not  underlined and $\underline c$ to all underlined parts of the partition. In both cases the tau-functions are
independent of all times $t_{3k}$.
\end{example}

\end{document}